\documentclass[%
 reprint,
superscriptaddress,
showpacs,
 amsmath,amssymb,
 aps,
prmaterials,
floatfix, longbibliography ]{revtex4-1}
\usepackage{graphicx}
\usepackage{dcolumn}
\usepackage{bm}
\usepackage{soul,color}
\usepackage[colorlinks,linkcolor=blue,anchorcolor=blue,citecolor=blue,urlcolor=blue]{hyperref}
\usepackage[]{datetime}
\begin{document}

\title{Flow of Polar and Nonpolar Liquids through Nanotubes:
    A Computational Study}


\author{Andrii Kyrylchuk}
\affiliation{Physics and Astronomy Department,
             Michigan State University,
             East Lansing, Michigan 48824-2320, USA}
\affiliation{Institute of Organic Chemistry,
             National Academy of Sciences of Ukraine,
             Murmanska Str.~5, 02660 Kyiv, Ukraine}

\author{David Tom\'anek}
\email[E-mail: ]{tomanek@msu.edu}
\affiliation{Physics and Astronomy Department,
             Michigan State University,
             East Lansing, Michigan 48824-2320, USA}

\date{\today}

\begin{abstract}
We perform {\em ab initio} density functional calculations to
study the flow of water, methanol and dimethyl ether through
nanotubes of carbon and boron nitride with different diameters and
chiralities. The liquids we choose are important solvents, with
water and methanol being polar and dimethyl ether being non-polar.
In terms of activation barriers for liquid transport, we find the
molecular-level drag to decrease with decreasing nanotube
diameter, but to be rather independent of the chiral index. We
also find molecules with higher polarity to experience higher drag
during the flow. Counter-intuitively, we find the drag for water
in boron nitride nanotubes not to exceed that in carbon nanotubes
due to frustration in competing long-range Coulomb interactions.
\end{abstract}



\maketitle



\section{Introduction}

The field of micro- and nanofluidics has been evolving rapidly
during the last decades, with its broad impact ranging from
medical devices~\cite{{Zhang2020},{Yue2019},{Bhatia2014}} to
advanced fabrics and energy conversion
devices~\cite{{Gravesen1993},{Whitesides2006},{Kohler2019}} and
advanced membranes for water desalination by reverse
osmosis~\cite{DT274}. Aligned carbon nanotube (CNT) membranes
attract much interest because of their well-defined pore size,
their capability to form composite matrices~\cite{Yang19} and
their resistance to biofouling~\cite{Rashid2017}. Many useful
properties of CNTs are also found in related BN nanotubes (BNNTs)
formed of hexagonal boron nitride~\cite{Chopra1995}. These
benefits combined with high thermal, chemical and mechanical
stability should make CNTs and BNNTs a very suitable material to
form membranes~\cite{Yang19}. Consequently, understanding the flow
of fluids inside nanometer-sized channels in these nanotubes is of
growing importance~\cite{Faucher2019}.

Since liquid flow inside nanotubes could not be observed with
sub-nanometer resolution so far, our understanding to date has
relied almost exclusively on molecular dynamics (MD) simulations.
Many force fields have been developed to describe the delicate
interplay among the strong and weak forces that determine the
behavior of liquids including water~\cite{%
{Kalra2003},{Zhu2003},{Thomas2009},{Hummer2001},{Kohler2019},%
    {Misra2017a}}, but none has been able to satisfactorily reproduce
all aspects of their behavior including their interaction with
solids in an unbiased manner. Such calculations provide valuable
information, but final results depend heavily on models used for
interatomic interactions, including the flexibility of bond
lengths and angles in water, molecular polarizability and
long-range electrostatic interactions~\cite{%
{Kohler2019},{Misra2017a}}. On the other hand, force fields based
on \textit{ab initio} total energy functionals including DFT are
nominally free of adjustable parameters, but high computational
requirements have severely limited their use.

Much insight has been obtained so far using different types of MD
simulations. Occurrence of water wires inside small CNTs has been
revealed by model force field simulations~\cite{%
{Kalra2003},{Zhu2003},{Thomas2009},{Hummer2001}}. There results
were confirmed by \textit{ab initio} density functional theory
(DFT) calculations~\cite{{Mann2003},{Dellago2003}} and Raman
spectroscopy observations~\cite{Cambre2010}. Formation of stacked
ring structures in wide CNTs with diameters ${\gtrsim}10$~{\AA}
has been corroborated by MD simulations based on model force
fields~\cite{{Kolesnikov2004},{Liu2005},{Koga2001},{Thomas2009}}
and DFT~\cite{Cicero2008}. These findings were experimentally
confirmed by IR spectroscopy~\cite{Byl2006} and X-ray
diffraction~\cite{Maniwa2005}. Still, there are very many open
questions~\cite{Faucher2019}.

There is some some experimental
evidence~\cite{%
{Holt2006a},{Eijkel2007},{Lauga2007},{Whitby2008},%
{Tsimpanogiannis2019}} that the flow velocity of liquids inside
nanometer-sized pores is several orders of magnitude higher than
what conventional theory based on Newtonian flow and the
Hagen-Poiseuille equation would predict~\cite{flow20}. There are
clearly some pitfalls in the way to translate and interpret
experimental data into microscopic fluid flow speeds~\cite{DT274}.
In particular, slip-flow of liquids inside nanochannels of
different types needs to be addressed. To date there are no
conclusive experimental or parameter-free \textit{ab initio}
results regarding the dependence of fluid flow inside nanotubes in
terms of nanotube type, diameter and chirality. Stull there are
many questions to be addressed regarding mass transport of liquids
through nanotubes and nanopores in general.

In this manuscript we address atomic-scale details of the flow of
water, methanol and dimethyl ether through nanotubes of carbon and
boron nitride with different diameters and chiralities. These
liquids are important solvents, with water and methanol being
polar and dimethyl ether being non-polar. We determine the
potential energy of isolated molecules along the inner nanotube
surface to determine activation barriers for diffusion, which
translate into molecular-level drag. We find this drag to decrease
with decreasing nanotube diameter, but to be rather independent of
the chiral index. We also find that molecules with higher polarity
experience higher drag during the flow.


\section{Computational approach}

Our computational approach to study liquid water and other
solvents interacting with carbon nanotubes is based on \textit{ab
    initio} density functional theory (DFT) as implemented in the
{\textsc{SIESTA}}~\cite{SIESTA} code. Unless specified otherwise,
we used the nonlocal Perdew-Burke-Ernzerhof (PBE)~\cite{PBE}
exchange-correlation functional, norm-conserving Troullier-Martins
pseudopotentials~\cite{Troullier91}, a double-$\zeta$ basis
including polarization orbitals, and a mesh cutoff energy of
$500$~Ry to determine the self-consistent charge density. We have
used periodic boundary conditions throughout the study, with
nanotube separated by ${\gtrsim}9$~{\AA} vacuum space.
Since the nanotube unit cells were long, we sampled the reciprocal
space by a uniform $1{\times}1{\times}2$ $k$-point
grid~\cite{Monkhorst-Pack76}. Systems with very large unit cells
have been represented by the $\Gamma$ point only. This provided us
with a precision in total energy of ${\lesssim}2$~meV/atom.
Geometries have been optimized using the conjugate gradient (CG)
method~\cite{Hestenes1952}, until none of the residual
Hellmann-Feynman forces exceeded $10^{-2}$~eV/{\AA}. While
computationally rather demanding, the DFT-PBE total energy
functional is free of adjustable parameters and has been used
extensively to provide an unbiased description water and its
interaction with solids~\cite{{Cicero2008},{Ambrosetti2011}}.
Selected MD simulations were performed using $0.3$~fs time steps,
which were sufficiently short to guarantee energy conservation in
a microcanonical ensemble.


\section{Results}

As mentioned above, the vast majority of atomistic MD simulations
is based on parameterized force fields, which offer hight degree
of numerical efficiency and allows to study the motion of several
thousand atoms simultaneously. The drawback of this approach is
its lack of universality and quantitative predictability: force
fields optimized for bulk fluids need to be changed at interfaces
and in situations, where long-range electrostatic interactions
play a role~\cite{{Kohler2019},{Misra2017a}}. For this reason, we
decided to use the \textit{ab initio} DFT formalism in our study.
In spite of its high computational demand, DFT is nominally free
of parameters and independent of predefined assumptions. This
approach has been validated in successfully predicting static and
dynamic properties of liquid water~\cite{DT274} and should provide
valuable information that should complement large-scale studies
with parameterized force fields.

\begin{figure}[t]
    \centering
    \includegraphics[width=0.8\columnwidth]{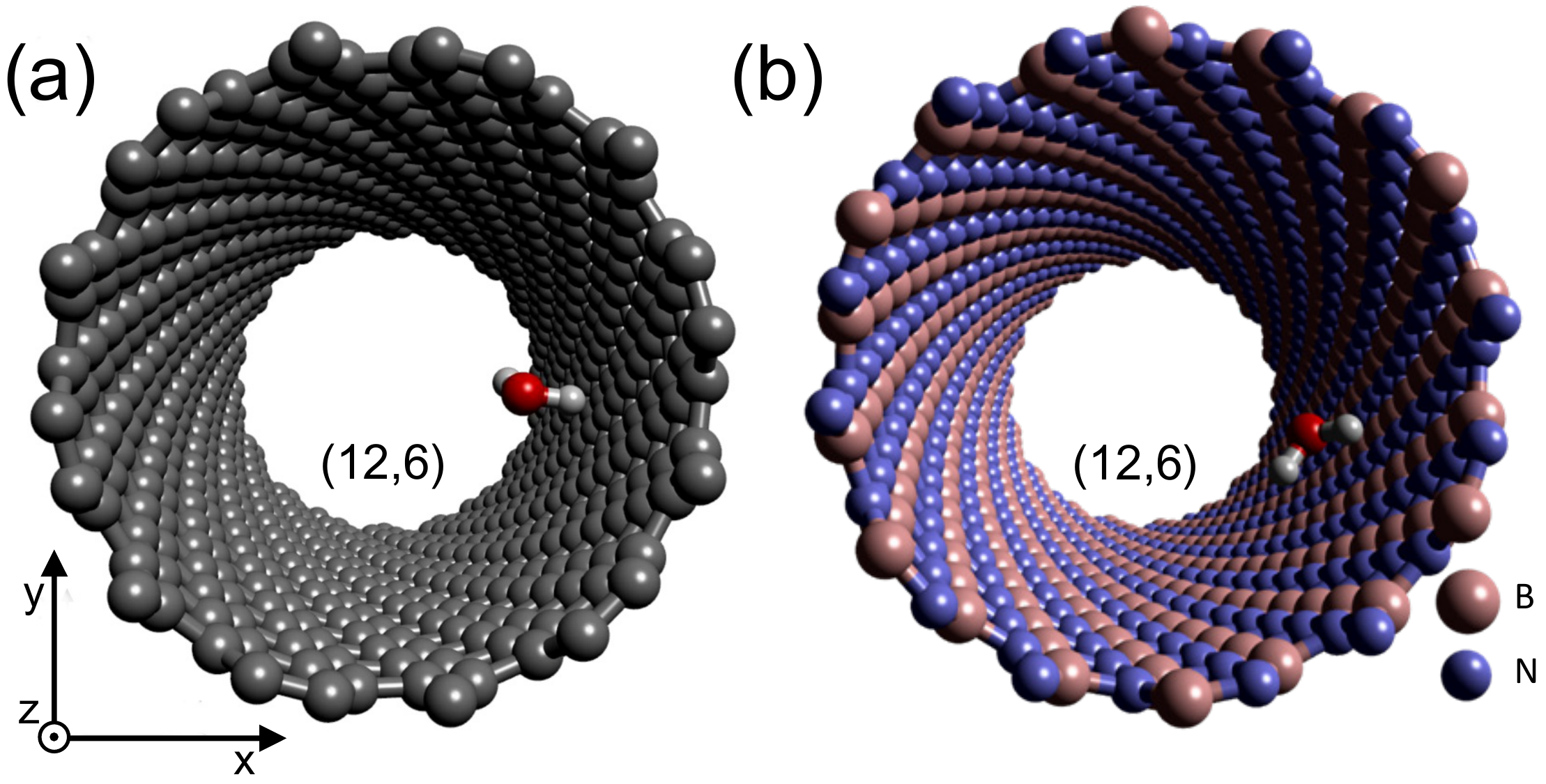}
    \caption{%
        A $(12,6)$ chiral nanotube of
        (a) carbon and (b) boron nitride (BN),
        containing one water molecule. }%
    \label{fig1}
\end{figure}

Since DFT poses serious limitations on the number of atoms in the
unit cell, small system sizes used in DFT calculations increase
the statistical noise and thus limit quantitative conclusions. To
avoid this problem, our following results are obtained using
static calculations rather than MD simulations. As a good
indication of the drag, which molecules experience at the
interface during flow through narrow cavities, we calculated the
potential energy of a single molecule drifting along nanotubes
with different composition, diameters, and chiral indices. These
results reveal potential energy barriers that should correlate
with the friction coefficients and slip lengths. We oriented the
nanotube axis along the $z$-direction and fixed all nanotube atoms
in their relaxed positions. We then determined the interaction
energy between the enclosed molecule $M$ and the nanotube, defined
by
\begin{equation}
    {\Delta}E = E_{tot}(M@NT) - E_{tot}(M) - E_{tot}(NT) \;.
\end{equation}
A negative value of ${\Delta}E$ indicates an energetic preference
for $M$ to be inside rather than to be isolated from the nanotube.
${\Delta}E(z)$ was calculated along the diffusion path of the
molecule by fixing only the $z$-coordinate of a representative
atom and relaxing all remaining molecular degrees of freedom,
starting with the molecule close to the inner nanotube surface.
This approach leaves the molecule free to find its optimum
position within the $x-y$ plane normal to the tube axis, to
optimize its shape and orientation. We considered a result to be
converged when the same final geometry with the same energy was
reached from different starting positions. The full ${\Delta}E(z)$
potential energy surface was obtained by a sequence of steps
involving displacements of the representative atom along the tube
axis. The `step size' was determined by the length of the nanotube
unit cell, which depends solely on the chiral index. In our
calculations we used step sizes ranging from
$0.2{\ldots}1.0$~{\AA}.


\subsection{Water interactions with CNTs}

\begin{figure*}[t]
    \centering
    \includegraphics[width=1.5\columnwidth]{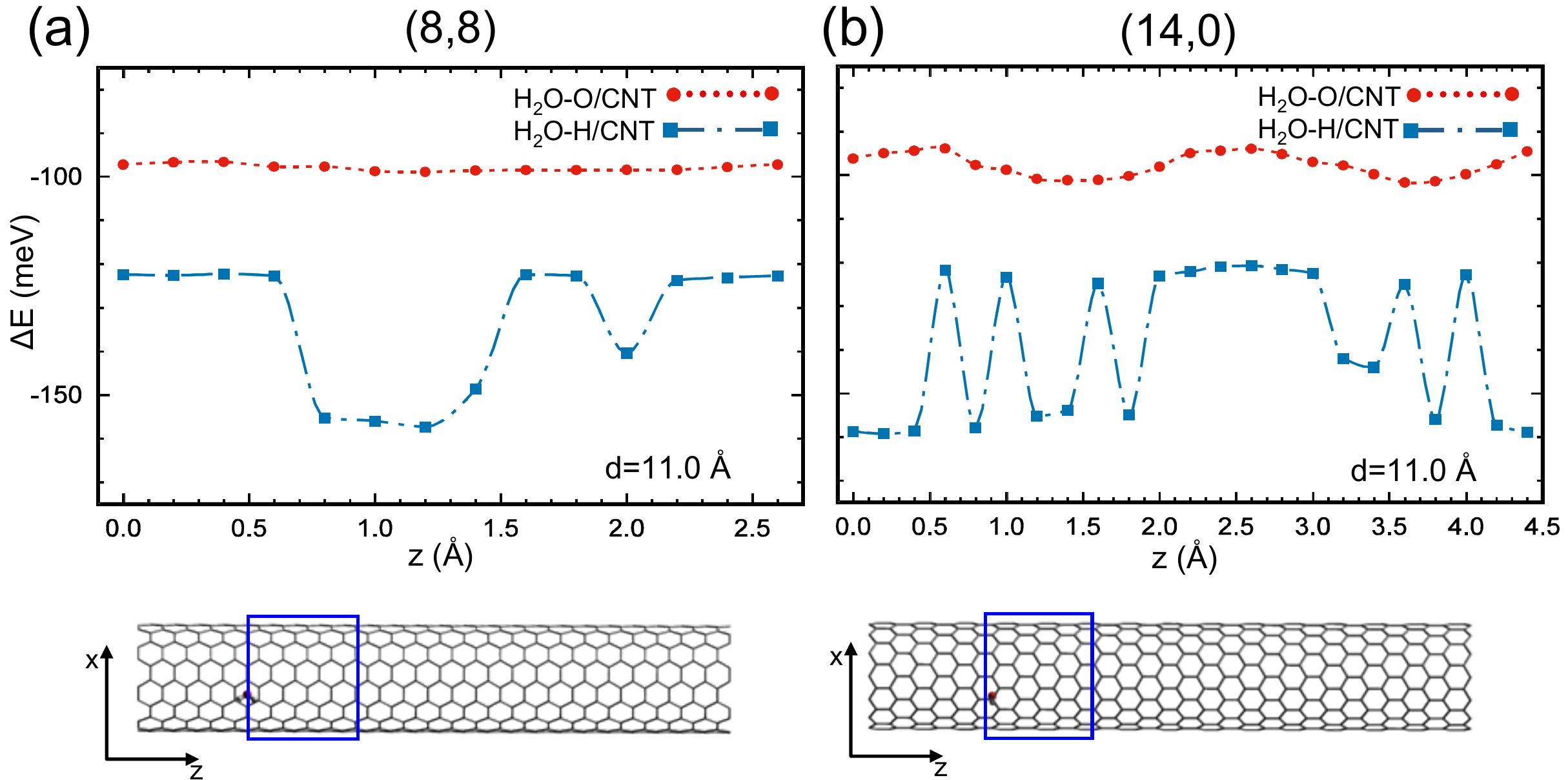}
    \caption{Interaction energy ${\Delta}E(z)$ between an isolated
        H$_2$O molecule and a surrounding %
        (a) armchair $(8,8)$ and %
        (b) zigzag $(14,0)$ CNT.
        The range of ${\Delta}E(z)$ values is the same in both panels.
        The $z$-coordinate represents the water position along the
        $z$-axis of the CNT. We distinguish H$_2$O-H/CNT configuration,
        where H-end of H$_2$O faces the wall, from H$_2$O-O/CNT, where the
        O-end faces the wall. Negative values of ${\Delta}E$ indicate
        energetic preference for H$_2$O entering the nanotube. The lower
        panels give a schematic view of the respective nanotubes.
        Nanotube diameters $d$ are indicated in the individual panels.
        The size of the unit cell considered is outlined by the
        blue box.
    }%
    \label{fig2}
\end{figure*}

Our first study was dedicated to the interaction of individual
water molecules with a CNT and a BNNT. The position of the
H$_{2}$O molecule along the tube axis, which is parallel to the
$z$-axis, is represented the $z-$coordinate of the oxygen atom.
`Snap shots' of a water molecule inside the $(12,6)$ CNT and BNNT
are shown in Fig.~\ref{fig1}.

Quantitative results for $E(z)$ for water inside an armchair
$(8,8)$ and a zigzag $(14,0)$ CNTs with essentially the same
diameter $d=11.0$~{\AA} are presented in Fig.~\ref{fig2}. We have
optimized the H$_{2}$O molecule using two starting configurations,
with the CNT inner wall facing either the H-atoms (H$_2$O-H/CNT)
or facing the oxygen (H$_2$O-O/CNT). Chemical intuition suggests a
repulsion between the oxygen lone electron pairs and the
$\pi$-system of the CNT, pushing the molecule toward the center.
The small positive charge on the hydrogen atoms, on the other
hand, should attract the molecule towards the wall. This is
confirmed by our results in Fig.~\ref{fig2}(a): A water molecule
gains ${\lesssim}98.9$~meV when entering an $(8,8)$ CNT with
oxygen facing the wall and as much as ${\approx}157.3$~meV when
the hydrogens face the wall. Of these two configurations, the one
with oxygen facing the wall is metastable. As seen in
Fig.~\ref{fig2}(b), we find roughly the same binding energies for
the two configurations in the $(14,0)$ CNT with the same diameter.

The weaker interaction of the metastable configuration with O
facing the wall also results in a very weak dependence of
${\Delta}E(z)$ on the position of the molecule both in the $(8,8)$
and in the $(14,0)$ CNT. In this orientation, H$_2$O may slip
along the CNT wall with essentially no barrier. The situation is
different in the stable configuration of water facing the CNT wall
with its hydrogen atoms. Besides a higher binding energy in this
orientation, also the barriers in ${\Delta}E(z)$ are significantly
higher, of the order of $40$~meV. The precise position of the
minima and maxima results from the hydrogen pair in H$_2$O finding
a match or a mismatch with C atoms in the graphitic CNT wall.
Especially in wide CNTs, the range of ${\Delta}E(z)$ values is
approximately the same independent of chirality, since it
essentially represents the interaction of a water molecule with a
graphene layer. It is only the sequence of minima and maxima that
distinguishes between chiral indices in the same way as different
sequences of maxima and minima occur when a water molecule crosses
a graphene sheet along different trajectories.

In the following investigations, we will only consider the stable
configuration of H$_2$O molecules facing the nanotube wall with
their hydrogen atoms.


\subsection{Water diffusion inside zigzag and armchair CNTs}

\begin{figure*}[t]
    \centering
    \includegraphics[width=1.5\columnwidth]{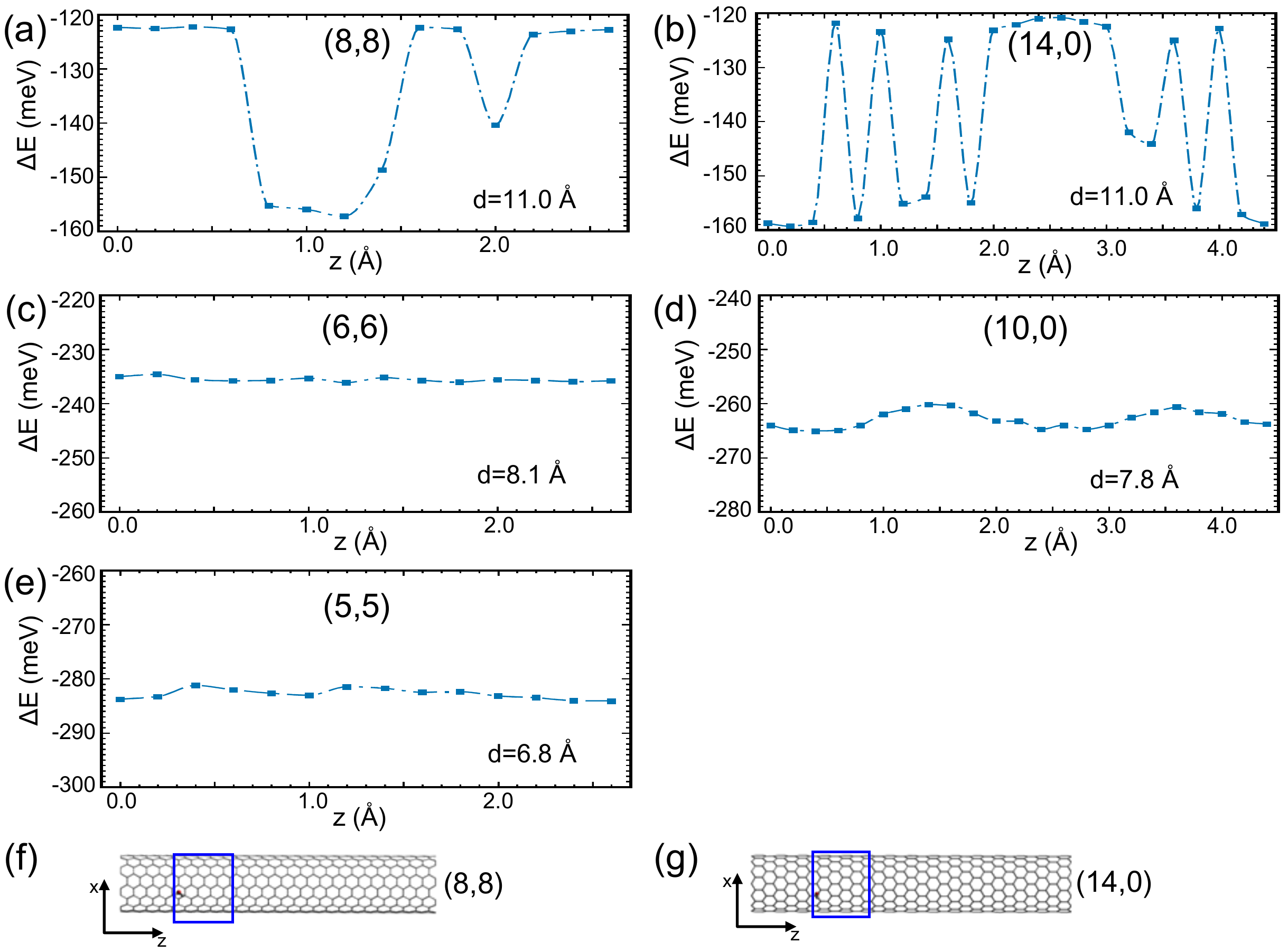}
    \caption{%
        Interaction energy ${\Delta}E(z)$ between an isolated H$_2$O
        molecule and a surrounding %
        (a) $(8,8)$, %
        (b) $(14,0)$, %
        (c) $(6,6)$, %
        (d) $(10,0)$, and %
        (e) $(5,5)$ CNT. %
        The $z$-coordinate represents the water position along the
        $z$-axis of the CNT. Only the stable water orientation
        with the hydrogens facing the wall are considered.
        The range of ${\Delta}E(z)$ values is $40$~meV in all panels.
        Schematic geometry is shown for %
        (f) the $(8,8)$ and %
        (g) the $(14,0)$ CNT, with the size of the unit
        cell considered being outlined by the blue box.
        Nanotube diameters $d$ are indicated in the individual panels.
    }%
    \label{fig3}
\end{figure*}

\begin{table}[b]
    \centering
    \caption{%
        Energy barriers $E_{p}$ for the diffusion of an isolated H$_2$O
        molecule along CNTs with a specific chiral index and diameter
        $d$. Reported results are for orientation of the molecule with
        its hydrogen pair facing the wall. %
    }%
    \begin{tabular}{ccc} %
        \hline \hline
        \textrm{Chiral Index} %
        & \textrm{$d$~({\AA})} %
        & \textrm{$E_{p}$~(meV) } %
        \\
        \hline%
        {(5,5)} %
        & {6.8} %
        & {2.8} %
        \\
        \hline%
        {(6,6)} %
        & {8.1} %
        & {1.5} %
        \\
        {(10,0)} %
        & {7.8} %
        & {4.9} %
        \\
        \hline%
        {(8,8)} %
        & {11.0} %
        & {35.0} %
        \\
        {(14,0)} %
        & {11.0} %
        & {38.5} %
        \\
        \hline%
        {(10,6)} %
        & {11.0} %
        & {37.6} %
        \\
        {(11,5)} %
        & {11.1} %
        & {36.5} %
        \\
        {(12,3)} %
        & {10.8} %
        & {37.3} %
        \\
        \hline \hline %
    \end{tabular}
    \label{table1}
\end{table}

We display the potential energy surface ${\Delta}E(z)$ for water
diffusion inside CNTs with selected chiral indices in
Fig.~\ref{fig3}. A more complete list of potential energy barriers
$E_p$ is presented in Table~\ref{table1}. As a general trend, we
notice that the size of the diffusion barriers decreases with
decreasing CNT diameter. We also find that the H$_2$O-CNT
interaction increases with decreasing nanotube diameter, making
narrower CNTs more hydrophilic. Thus, with decreasing tube
diameter $d$, CNTs turn from hydrophobic ($d{\rightarrow}\infty$
for graphene) to increasingly hydrophilic and eventually to
hydrophobic at diameters too small to accommodate a water
molecule. The increase of the H$_2$O-CNT interaction and reduction
of activation barriers is linked to the fact that in very narrow
nanotubes, atoms all around the nanotube perimeter interact with
the enclosed molecule. While stabilizing the enclosed molecule, it
results in frustrated geometries that reduce the dependence of the
interaction energy ${\Delta}E$ on the position of the molecule
$z$. This general trend has been identified in published MD
simulations based on parameterized force
fields~\cite{{Kannam2013},{Falk2010}}.
In agreement with our findings, these studies found that slip
lengths of water molecules decreased with increasing CNT diameter
and asymptotically approached the value for water on planar
graphene.

The presented quantitative findings compare well with published
DFT-PBE results for the drift of a monolayer of 2D ice on
graphene, suggesting energy barriers of ${\approx}15$~meV/H$_2$O
along the zigzag and ${\approx}25$~meV/H$_2$O along the armchair
direction~\cite{Boukhvalov2013}. We note that the strong
H$_2$O-H$_2$O interaction in ice does not alow individual water
molecules the same configurational freedom as the geometry
considered here.

Our results in Table~\ref{table1} also suggest that potential
energy barriers $E_p$ in zigzag CNTs are typically higher by
${\approx}3$~meV than in armchair nanotubes. At this point, we
should realize that drift occurs along the armchair direction in
CNTs with a zigzag edge and along the zigzag direction in
nanotubes with an armchair edge. Even though the small difference
between activation barriers approaches the precision limit of
PBE-DFT, we are pleased that our results for a single H$_2$O
molecule match those for 2D ice/graphene~\cite{Boukhvalov2013}
quite well. A similar small difference between activation barriers
in zigzag and armchair nanotubes, based on MD simulations with
parameterized force fields, has been reported
previously~\cite{{Wei2018},{Sam2019}}.

We noted another point of interest when studying the passage of a
water molecule through a very narrow $(5,5)$ nanotube. Considering
the van der Waals diameter $d_{vdW}$(H$_2$O)$=2.8$~{\AA} of a
water molecule~\cite{Jorgensen83} and the van der Waals radius
$r_{vdW}$(C$_{atom}$)$=1.8$~{\AA}~\cite{Bondi1964} of a carbon
atom, the sum $d_{vdW}$(H$_2$O)$+2r_{vdW}$(C$_{atom}$)$=6.4$~{\AA}
is only $0.5$~{\AA} smaller than the diameter $d=6.9$~{\AA} of the
$(5,5)$ CNT. In other words, this is a rather tight fit for the
enclosed molecule. Na\"{\i}vely, pressing the molecule against the
wall should increase the activation barrier for diffusion. In
reality, the opposite is true. The molecule now interacts with
many atoms along the perimeter of the surrounding nanotube, with
unfavorable interactions compensating favorable interactions in
different regions. The resulting frustration lowers the activation
barriers for diffusion.


\subsection{Water diffusion inside chiral CNTs}

\begin{figure*}[t]
    \centering
    \includegraphics[width=1.5\columnwidth]{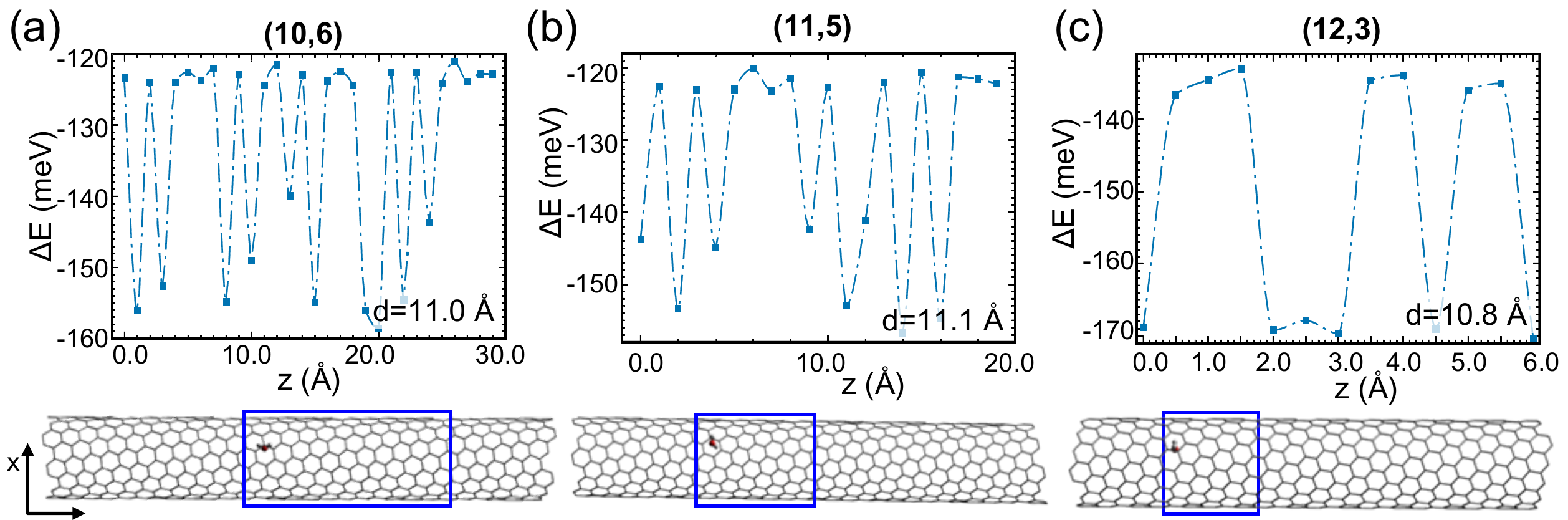}
    \caption{%
        Interaction energy ${\Delta}E(z)$ between an isolated H$_2$O
        molecule and a surrounding %
        (a) $(10,6)$, %
        (b) $(11,5)$ and %
        (c) $(12,3)$ CNT. %
        The range of ${\Delta}E(z)$ values is $40$~meV in all panels.
        Nanotube diameters $d$ are indicated in the individual panels.
        The $z$-coordinate represents the water position along the
        $z$-axis of the CNT. Only the stable water orientation
        with the hydrogens facing the wall are considered. %
        Schematic geometries including the unit cells considered
        are shown below the respective ${\Delta}E(z)$ plots.
    }%
    \label{fig4}
\end{figure*}

To complete our study of water in CNTs, we present our results for
water inside three chiral nanotubes with a similar diameter
$d{\approx}11$~{\AA} in Fig.~\ref{fig4}. We used different step
sizes due to the different unit cells, which depend on the chiral
index. Our findings of essentially the same energy barrier values
$E_p{\approx}30$~meV fall in line with all our other results,
suggesting that the energy barriers depend primarily in the
nanotube diameter, with only minor dependence on the chiral index.

We have also filled the chiral $(12,6)$ CNT with water and used a
DFT-PBE based MD simulation to study the flow driven by pressure
difference between the tube ends, which is counter-balanced by
drag. As seen in Video~\ref{video1}, the chirality of the
surrounding CNT does not exert a sufficient torque on the water
column to induce axial rotation.

\begin{video}[b]
\includegraphics[width=0.9\columnwidth]{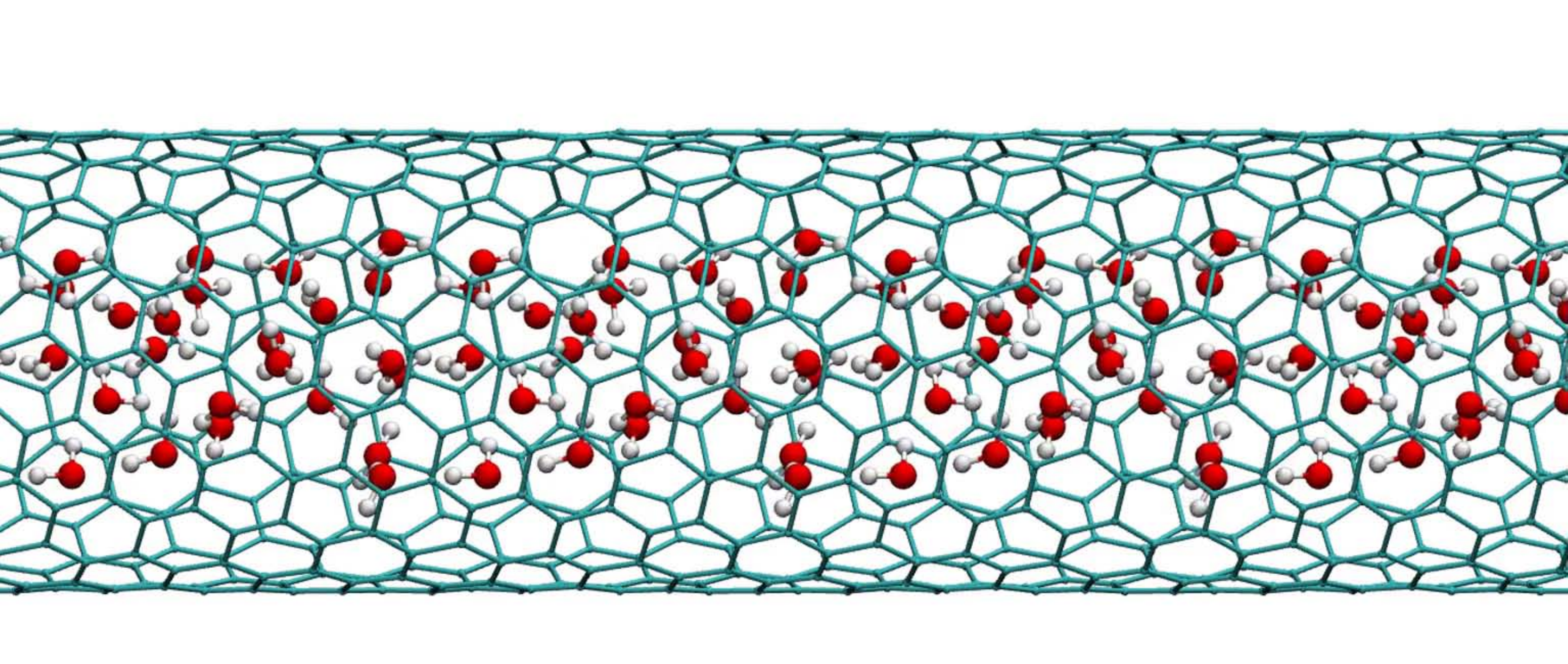}
\setfloatlink{video1.mp4} %
\caption{MD simulation of water flow through a chiral $(12,6)$
CNT.}
\label{video1}
\end{video}


\subsection{Water diffusion inside BNNTs}

\begin{figure*}[t]
    \centering
    \includegraphics[width=1.5\columnwidth]{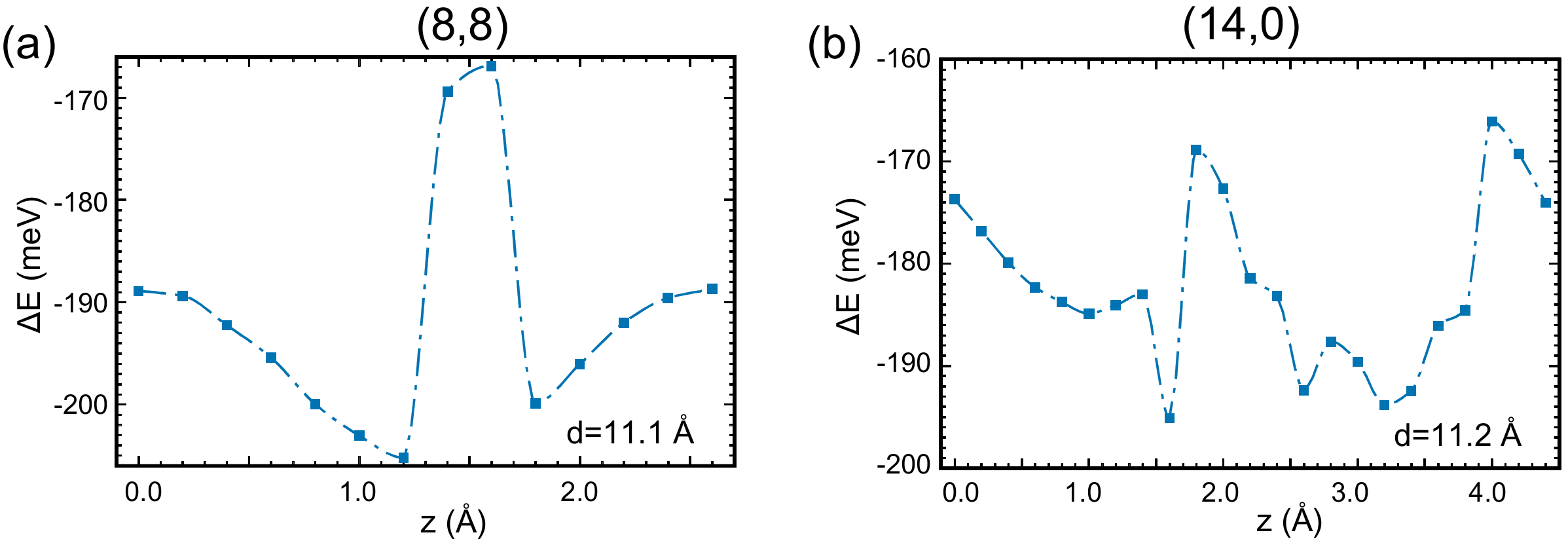}
    \caption{%
        Interaction energy ${\Delta}E(z)$ between an isolated water
        molecule contained in an %
        (a) $(8,8)$ and %
        (b) $(14,0)$ %
        BNNT. Nanotube diameters $d$ are indicated in the individual
        panels. The range of ${\Delta}E(z)$ values is $40$~meV
        in all panels. The $z$-coordinate represents the position
        of the molecule along the $z$-axis of the BNNT.
    }%
    \label{fig5}
\end{figure*}

Similar to graphite, hexagonal boron nitride ($h$-BN) is a layered
material capable of forming nanotubes~\cite{Chopra1995}. Even
though $h$-BN is isoelectronic to graphene, B-N bonds are slightly
longer than C-C bonds in graphene. There is a corresponding
${\approx}2\%$ increase in the diameter of BNNTs over CNTs with
the same chiral index. Unlike CNTs, all BNNTs are wide-gap
insulators. The polar nature of the B-N bond provides additional
attraction to polar molecules such as water. In analogy to our CNT
results, we present the interaction energy of a water molecule
with surrounding $(8,8)$ and $(14,0)$
BNNTs in Fig.~\ref{fig5}.

Unlike in CNTs of similar diameter, we found optimization of water
inside BNNTs to be more demanding, likely due to the long-range
nature of the Coulomb interactions dominating there. While we
could not guarantee identifying the lowest energy state among many
optima with similar energies, our results in Fig.~\ref{fig5}
provide consistent trends. For the sake of comparison with the
corresponding CNTs, we summed up the van der Waals radii
$r_{vdW}($N$)=1.55$~{\AA} of nitrogen~\cite{Bondi1964},
$r_{vdW}($B$)=1.92$~{\AA} of boron~\cite{Mantina2009}, and the van
der Waals diameter $d_{vdW}$(H$_2$O)$=2.8$~{\AA} of a water
molecule~\cite{Jorgensen83}. The corresponding diameter of a BNNT
that would contain water in a tight fit is $6.3$~{\AA}.
This value is significantly smaller than the diameter
$d{\approx}11$~{\AA} of $(8,8)$ and $(14,0)$ BNNTs, indicating
that -- similar to CNTs -- the water molecule is not sterically
constrained in these nanotubes.

The suggested additional attraction of water to walls of BNNTs in
comparison to CNTs is best illustrated by comparing ${\Delta}E$
values for H$_2$O in $(8,8)$ and $(14,0)$ nanotubes of carbon and
BN presented in Figs.~\ref{fig3}(a-b) and \ref{fig5}. For both
chiral indices, we see a significant stabilization of water
molecules in BNNTs by ${\approx}40$~meV over CNTs. A na\"{\i}ve
expectation of corresponding increase of activation barriers for
diffusion has not materialized; we find a similar value
$E_p{\approx}40$~meV in all these nanotubes.

The similarity of the activation barriers for water in BNNTs and
CNTs has an interesting physical origin. Net charges do not play a
major role in the non-polar bond between H$_2$O and a CNT, with
hydrogens facing the wall, which is rather local. The situation is
rather different for water inside a BNNT. The Coulomb interaction
between the polar H$_2$O molecule is long-ranged, involving
charges on B and N sites all around the BNNT diameter. Whereas the
water dipole encounters a net stabilization inside a BNNT, the
competition between significant attractive and repulsive forces
that are long-ranged leads to a degree of frustration that
decreases the dependence of ${\Delta}E$ on the axial position of
the molecule.

Controversial results have been reported on water flow through
CNTs and BNNTs. Results claiming a faster flow of water in BNNTs
than in CNTs~\cite{{Aluru2007},{Suk2008}} directly contradict
results to the opposite~\cite{{Ritos2014},{Wei2018},{Secchi2016}}.

Setting aside that additional issue of different entry barriers
for water in CNTs and BNNTs~\cite{Thomas2014}, diffusion barriers
for drifting water, which depend on local orbital hybridization
with nanotube atoms as well as long-range Coulomb forces, can not
be described adequately by parameterized short-range
potentials~\cite{{Hilder2010},{Melillo2011}}, which tend to
exaggerate the value of $E_p$ in BNNTs.

Unlike inside a cylindrical BNNT, there is much less frustration
in terms of competing attractive and repulsive forces on planar
$h$-BN. As evidenced by {\em ab initio} DFT
calculations~\cite{Tocci2014}, this leads to an increased
modulation of ${\Delta}E$ across the $h$-BN surface in comparison
to graphene.


\subsection{Diffusion of different molecules inside CNTs}

The above-mentioned application of CNTs for the filtration of
water can be trivially extended to other liquids, and there is
some experimental and theoretical evidence in the
direction~\cite{%
    {Faucher2019},{Gelb1999},{Czwartos2005},{Srivastava2004},%
    {Whitby2008}}. Besides water, we studied methanol (CH$_3$-OH,
Me-OH) and dimethyl ether (CH$_3$-O-CH$_3$, MeOMe) molecules that
are comparable in size. All three substances are important
solvents, with the dielectric constant decreasing to H$_2$O,
Me-OH, and to MeOMe, as represented in Table~\ref{table2}. We
studied the potential energy surface ${\Delta}E(z)$ for each of
these molecules drifting along the $(8,8)$ CNT in much the same
way as described for water earlier. Our results for ${\Delta}E(z)$
are displayed in Fig.~\ref{fig5} and for the energy barriers $E_p$
in Table~\ref{table2}.

\begin{figure*}[t]
    \centering
    \includegraphics[width=1.5\columnwidth]{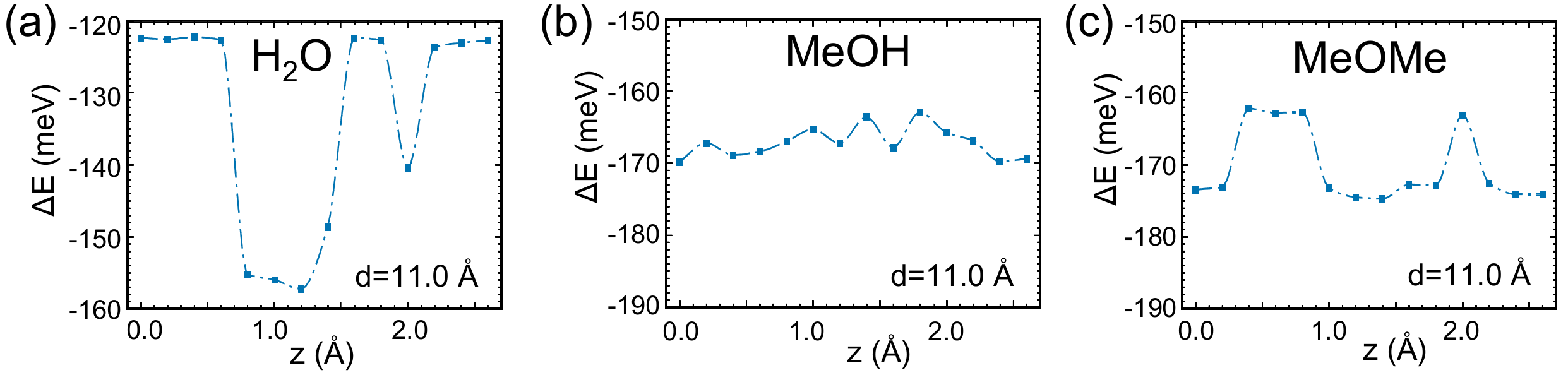}
    \caption{%
        Interaction energy ${\Delta}E(z)$ between an isolated %
        (a) water (H$_2$O), %
        (b) methanol (CH$_3$-OH, MeOH), and %
        (c) dimethyl ether (CH$_3$-O-CH$_3$, MeOMe) %
        molecule and a surrounding $(8,8)$ CNT. %
        The range of ${\Delta}E(z)$ values is $40$~meV and
        nanotube diameters $d$ are indicated in all panels.
        The $z$-coordinate represents the position of the molecule along the
        $z$-axis of the CNT.
    }%
    \label{fig6}
\end{figure*}

Sequential substitution of hydrogen atoms by methyl groups
substantially lowers the dipole moment $p$ of the molecule and the
dielectric constant $\epsilon$ of the liquid. This lowers the
interaction of the molecule with the inner CNT wall and lower
diffusion barriers, as seen clearly by comparing our results for
H$_2$O and methanol in Fig.~\ref{fig6}. As seen in
Fig.~\ref{fig6}(c), this trend is not completely followed by
dimethyl ether with a very low static dipole moment.

While increasing the number of hydrogens substituted by methyl
groups lowers the dipole moment, the concurrent growth of
molecular size increases the molecular polarizability $\alpha$,
reaching its maximum in dimethyl ether. It may be argued that
higher values of $\alpha$ correlate with higher reactivity and
stronger interaction with the tube wall. In comparison to water,
we see a similar bonding enhancement, indicated by lower values of
${\Delta}E$, in both methanol and dimethyl ether. Our results
suggest that increase in either $p$ or $\alpha$ tend to increase
the activation barriers for diffusion, which are largest for water
and dimethyl ether. We should note that changes in the barrier
size are rather modest in the solvents we consider.

\begin{table}[h]
    \centering
    \caption{%
        Bulk dielectric constant $\epsilon$, molecular polarizability $\alpha$
        and energy barrier $E_{p}$ for the diffusion of an isolated solvent molecule
        inside an $(8,8)$ armchair CNT. %
    }%
    \begin{tabular}{lccc} %
        \hline \hline
        \textrm{Solvent} %
        & \textrm{Dielectric} %
        & \textrm{Polarizability $\alpha$~\cite{H2O-CRC86}} 
        & \textrm{$E_{p}$} %
        \\
        \textrm{} %
        & \textrm{constant $\epsilon$} %
        & \textrm{(${\times}10^{-24}$~cm$^{3}$)} 
        & \textrm{(meV)} %
        \\
        \hline%
        {Water} %
        & {$80.1$} %
        & {$1.45$} %
        & {$35.0$} %
        \\
        {Methanol} %
        & {$33.0$} %
        & {$3.23{\ldots}3.32$} %
        & {$6.9$} %
        \\
        {Dimethyl Ether} %
        & {6.2} %
        & {$5.16\ldots5.84$} %
        & {$12.6$} %
        \\
        \hline \hline %
    \end{tabular}
    \label{table2}
\end{table}

\section{Discussion}

So far, most of our calculations focussed on a single molecule
drifting along a CNT or a BNNT. Even though this geometry is
different from a nanotube filled with many molecules representing
a liquid, a relation between molecular diffusion barriers and the
resulting drag force in the liquid has been identified
previously~\cite{Wei2018}. The specific claim, based on MD
simulation results for $(8,8)$ and $(14,0)$ CNTs that contain a
single H$_2$O molecule or are filled with water, is that the
trends in the friction coefficients are the same in the two
cases~\cite{Wei2018}. Similar to our findings, consistently higher
friction coefficients were reported~\cite{Wei2018} in the zigzag
$(14,0)$ CNT than in the armchair $(8,8)$ CNT, which contained
either a single H$_2$O molecule or were filled with water. The
friction coefficients in the different CNTs were very close in the
single-molecule case and increased in size with increasing water
filling level.

We believe that the latter finding is closely related to the way,
in which activation barriers for diffusion are determined. The
potential energy surfaces ${\Delta}E(z)$ for different molecules
and nanotubes, which we present in our study, are the result of a
large number of static constrained structure optimizations. MD
simulations of an isolated molecule drifting freely within the
nanotube void, on the other hand, will offer a different picture.
Even though the potential energy of such a drifting molecule does
change along its trajectory, the molecule will unlikely encounter
all maxima and minima in the potential energy surface. This would
be particulary true for vacancy defects that may not hinder the
axial drift of a molecule at finite velocity~\cite{Joseph2008}.
Consequently, the spread of potential energies encountered by a
freely drifting molecule will be lower than that found in the
${\Delta}E(z)$ surfaces presented here.

Molecules in a nanotube completely filled with a liquid are
radially constrained by the low compressibility of the liquid and
the rigidity of the enclosing CNT wall. In comparison to a freely
drifting molecule that barely touches the inner walls, molecules
at the interface between the liquid column and the nanotube are
forced to `hug' the surface closely, thus probing the potential
profile of the nanotube wall more intimately. In general, the
liquid column propagates in a plug-like motion along the
nanotube~\cite{Hanasaki2006}. Its viscosity -- unlike the friction
coefficient -- does not contribute much to the friction
coefficient~\cite{Joly2011}. To a small degree, the flexibility of
the CNT walls has been shown to further reduce the friction
coefficient in comparison to a rigid
wall~\cite{{Sam2017},{Tao2018}}.

The calculated relative barriers for the passage of water and
other solvents show that the CNT membranes can indeed be used for
the separation of liquids and water purification including
desalination~\cite{{DT274},{Yang19}}. Viable applications of
aligned CNT membranes may then extend from removal of organic
components from water to the purification of gasoline and other
petroleum fractions~\cite{Srivastava2004}. The former is an
especially pressing issue, since state-of-the-art polymer
membranes used in the reverse osmosis process are easily destroyed
by hydrocarbons~\cite{Jassby2018}. Polymer membranes also have
insufficient selectivity to non-charged organic
molecules~\cite{Albergamo2018} and are prone to
bio-fouling~\cite{{Werber2016},{Werber2016a},{Elimelech2011}}.



\section{Summary and Conclusions}

We have conducted {\em ab initio} DFT studies of the drift of
water, methanol and dimethyl ether molecules inside nanotubes of
carbon and boron nitride with different diameters and chiral
indices. The liquids we choose are important solvents, with water
and methanol being polar and dimethyl ether being non-polar. In
terms of activation barriers for transport, we find the
molecular-level drag to decrease with decreasing nanotube
diameter, but to be rather independent of the chiral index. We
also found molecules with higher polarity or polarization to
experience higher drag during the flow. Rather
counter-intuitively, we found the drag for water molecules in
boron nitride molecules not exceed that in carbon nanotubes due to
frustration in competing long-range Coulomb interactions. We
expect that the trends identified in this study may help to design
nanotube membranes for the filtration of water and separation of
solvents.


\begin{acknowledgments}
D.T. acknowledges financial support by the NSF/AFOSR EFRI 2-DARE
grant number EFMA-1433459. A.K. acknowledges financial support by
the Fulbright program. We thank Aleksandr Noy for valuable
discussions. Computational resources have been provided by the
Michigan State University High Performance Computing Center.
\end{acknowledgments}


%


\begin{thebibliography}{0}%
\makeatletter
\providecommand \@ifxundefined [1]{%
 \@ifx{#1\undefined}
}%
\providecommand \@ifnum [1]{%
 \ifnum #1\expandafter \@firstoftwo
 \else \expandafter \@secondoftwo
 \fi
}%
\providecommand \@ifx [1]{%
 \ifx #1\expandafter \@firstoftwo
 \else \expandafter \@secondoftwo
 \fi
}%
\providecommand \natexlab [1]{#1}%
\providecommand \enquote  [1]{``#1''}%
\providecommand \bibnamefont  [1]{#1}%
\providecommand \bibfnamefont [1]{#1}%
\providecommand \citenamefont [1]{#1}%
\providecommand \href@noop [0]{\@secondoftwo}%
\providecommand \href [0]{\begingroup \@sanitize@url \@href}%
\providecommand \@href[1]{\@@startlink{#1}\@@href}%
\providecommand \@@href[1]{\endgroup#1\@@endlink}%
\providecommand \@sanitize@url [0]{\catcode `\\12\catcode `\$12\catcode
  `\&12\catcode `\#12\catcode `\^12\catcode `\_12\catcode `\%12\relax}%
\providecommand \@@startlink[1]{}%
\providecommand \@@endlink[0]{}%
\providecommand \url  [0]{\begingroup\@sanitize@url \@url }%
\providecommand \@url [1]{\endgroup\@href {#1}{\urlprefix }}%
\providecommand \urlprefix  [0]{URL }%
\providecommand \Eprint [0]{\href }%
\providecommand \doibase [0]{http://dx.doi.org/}%
\providecommand \selectlanguage [0]{\@gobble}%
\providecommand \bibinfo  [0]{\@secondoftwo}%
\providecommand \bibfield  [0]{\@secondoftwo}%
\providecommand \translation [1]{[#1]}%
\providecommand \BibitemOpen [0]{}%
\providecommand \bibitemStop [0]{}%
\providecommand \bibitemNoStop [0]{.\EOS\space}%
\providecommand \EOS [0]{\spacefactor3000\relax}%
\providecommand \BibitemShut  [1]{\csname bibitem#1\endcsname}%
\let\auto@bib@innerbib\@empty
\end{thebibliography}%


\begin{thebibliography}{67}%
\makeatletter
\providecommand \@ifxundefined [1]{%
 \@ifx{#1\undefined}
}%
\providecommand \@ifnum [1]{%
 \ifnum #1\expandafter \@firstoftwo
 \else \expandafter \@secondoftwo
 \fi
}%
\providecommand \@ifx [1]{%
 \ifx #1\expandafter \@firstoftwo
 \else \expandafter \@secondoftwo
 \fi
}%
\providecommand \natexlab [1]{#1}%
\providecommand \enquote  [1]{``#1''}%
\providecommand \bibnamefont  [1]{#1}%
\providecommand \bibfnamefont [1]{#1}%
\providecommand \citenamefont [1]{#1}%
\providecommand \href@noop [0]{\@secondoftwo}%
\providecommand \href [0]{\begingroup \@sanitize@url \@href}%
\providecommand \@href[1]{\@@startlink{#1}\@@href}%
\providecommand \@@href[1]{\endgroup#1\@@endlink}%
\providecommand \@sanitize@url [0]{\catcode `\\12\catcode
`\$12\catcode
  `\&12\catcode `\#12\catcode `\^12\catcode `\_12\catcode `\%12\relax}%
\providecommand \@@startlink[1]{}%
\providecommand \@@endlink[0]{}%
\providecommand \url  [0]{\begingroup\@sanitize@url \@url }%
\providecommand \@url [1]{\endgroup\@href {#1}{\urlprefix }}%
\providecommand \urlprefix  [0]{URL }%
\providecommand \Eprint [0]{\href }%
\providecommand \doibase [0]{http://dx.doi.org/}%
\providecommand \selectlanguage [0]{\@gobble}%
\providecommand \bibinfo  [0]{\@secondoftwo}%
\providecommand \bibfield  [0]{\@secondoftwo}%
\providecommand \translation [1]{[#1]}%
\providecommand \BibitemOpen [0]{}%
\providecommand \bibitemStop [0]{}%
\providecommand \bibitemNoStop [0]{.\EOS\space}%
\providecommand \EOS [0]{\spacefactor3000\relax}%
\providecommand \BibitemShut  [1]{\csname bibitem#1\endcsname}%
\let\auto@bib@innerbib\@empty
\bibitem [{\citenamefont {Zhang}\ \emph {et~al.}(2020)\citenamefont {Zhang},
  \citenamefont {Huang}, \citenamefont {Qian}, \citenamefont {Chen},
  \citenamefont {Wen},\ and\ \citenamefont {Jiang}}]{Zhang2020}%
  \BibitemOpen
  \bibfield  {author} {\bibinfo {author} {\bibfnamefont {Zhen}\ \bibnamefont
  {Zhang}}, \bibinfo {author} {\bibfnamefont {Xiaodong}\ \bibnamefont {Huang}},
  \bibinfo {author} {\bibfnamefont {Yongchao}\ \bibnamefont {Qian}}, \bibinfo
  {author} {\bibfnamefont {Weipeng}\ \bibnamefont {Chen}}, \bibinfo {author}
  {\bibfnamefont {Liping}\ \bibnamefont {Wen}}, \ and\ \bibinfo {author}
  {\bibfnamefont {Lei}\ \bibnamefont {Jiang}},\ }\bibfield  {title} {\enquote
  {\bibinfo {title} {Engineering smart nanofluidic systems for artificial ion
  channels and ion pumps: From single-pore to multichannel membranes},}\ }\href
  {\doibase 10.1002/adma.201904351} {\bibfield  {journal} {\bibinfo  {journal}
  {Adv. Mater.}\ }\textbf {\bibinfo {volume} {32}},\ \bibinfo {pages} {1904351}
  (\bibinfo {year} {2020})}\BibitemShut {NoStop}%
\bibitem [{\citenamefont {Yue}\ \emph {et~al.}(2019)\citenamefont {Yue},
  \citenamefont {Tan}, \citenamefont {Li}, \citenamefont {Liu},\ and\
  \citenamefont {Wang}}]{Yue2019}%
  \BibitemOpen
  \bibfield  {author} {\bibinfo {author} {\bibfnamefont {Wan-Qing}\
  \bibnamefont {Yue}}, \bibinfo {author} {\bibfnamefont {Zheng}\ \bibnamefont
  {Tan}}, \bibinfo {author} {\bibfnamefont {Xiu-Ping}\ \bibnamefont {Li}},
  \bibinfo {author} {\bibfnamefont {Fei-Fei}\ \bibnamefont {Liu}}, \ and\
  \bibinfo {author} {\bibfnamefont {Chen}\ \bibnamefont {Wang}},\ }\bibfield
  {title} {\enquote {\bibinfo {title} {Micro/nanofluidic technologies for
  efficient isolation and detection of circulating tumor cells},}\ }\href
  {\doibase 10.1016/j.trac.2019.06.009} {\bibfield  {journal} {\bibinfo
  {journal} {TrAC, Trends Anal. Chem.}\ }\textbf {\bibinfo {volume} {117}},\
  \bibinfo {pages} {101--115} (\bibinfo {year} {2019})}\BibitemShut {NoStop}%
\bibitem [{\citenamefont {Bhatia}\ and\ \citenamefont
  {Ingber}(2014)}]{Bhatia2014}%
  \BibitemOpen
  \bibfield  {author} {\bibinfo {author} {\bibfnamefont {Sangeeta~N.}\
  \bibnamefont {Bhatia}}\ and\ \bibinfo {author} {\bibfnamefont {Donald~E.}\
  \bibnamefont {Ingber}},\ }\bibfield  {title} {\enquote {\bibinfo {title}
  {Microfluidic organs-on-chips},}\ }\href {\doibase 10.1038/nbt.2989}
  {\bibfield  {journal} {\bibinfo  {journal} {Nat. Biotechnol.}\ }\textbf
  {\bibinfo {volume} {32}},\ \bibinfo {pages} {760--772} (\bibinfo {year}
  {2014})}\BibitemShut {NoStop}%
\bibitem [{\citenamefont {Gravesen}\ \emph {et~al.}(1993)\citenamefont
  {Gravesen}, \citenamefont {Branebjerg},\ and\ \citenamefont
  {Jensen}}]{Gravesen1993}%
  \BibitemOpen
  \bibfield  {author} {\bibinfo {author} {\bibfnamefont {P}~\bibnamefont
  {Gravesen}}, \bibinfo {author} {\bibfnamefont {J}~\bibnamefont {Branebjerg}},
  \ and\ \bibinfo {author} {\bibfnamefont {O~S}\ \bibnamefont {Jensen}},\
  }\bibfield  {title} {\enquote {\bibinfo {title} {Microfluidics-a review},}\
  }\href {\doibase 10.1088/0960-1317/3/4/002} {\bibfield  {journal} {\bibinfo
  {journal} {J. Micromech. Microeng.}\ }\textbf {\bibinfo {volume} {3}},\
  \bibinfo {pages} {168--182} (\bibinfo {year} {1993})}\BibitemShut {NoStop}%
\bibitem [{\citenamefont {Whitesides}(2006)}]{Whitesides2006}%
  \BibitemOpen
  \bibfield  {author} {\bibinfo {author} {\bibfnamefont {George~M.}\
  \bibnamefont {Whitesides}},\ }\bibfield  {title} {\enquote {\bibinfo {title}
  {The origins and the future of microfluidics},}\ }\href {\doibase
  10.1038/nature05058} {\bibfield  {journal} {\bibinfo  {journal} {Nature}\
  }\textbf {\bibinfo {volume} {442}},\ \bibinfo {pages} {368--373} (\bibinfo
  {year} {2006})}\BibitemShut {NoStop}%
\bibitem [{\citenamefont {K{\"{o}}hler}\ \emph {et~al.}(2019)\citenamefont
  {K{\"{o}}hler}, \citenamefont {Bordin}, \citenamefont {de~Matos},\ and\
  \citenamefont {Barbosa}}]{Kohler2019}%
  \BibitemOpen
  \bibfield  {author} {\bibinfo {author} {\bibfnamefont {Mateus~H.}\
  \bibnamefont {K{\"{o}}hler}}, \bibinfo {author} {\bibfnamefont
  {Jos{\'{e}}~R.}\ \bibnamefont {Bordin}}, \bibinfo {author} {\bibfnamefont
  {Carolina~F.}\ \bibnamefont {de~Matos}}, \ and\ \bibinfo {author}
  {\bibfnamefont {Marcia~C.}\ \bibnamefont {Barbosa}},\ }\bibfield  {title}
  {\enquote {\bibinfo {title} {Water in nanotubes: The surface effect},}\
  }\href {\doibase 10.1016/j.ces.2019.03.062} {\bibfield  {journal} {\bibinfo
  {journal} {Chem. Eng. Sci.}\ }\textbf {\bibinfo {volume} {203}},\ \bibinfo
  {pages} {54--67} (\bibinfo {year} {2019})}\BibitemShut {NoStop}%
\bibitem [{\citenamefont {Tom\'anek}\ and\ \citenamefont
  {Kyrylchuk}(2019)}]{DT274}%
  \BibitemOpen
  \bibfield  {author} {\bibinfo {author} {\bibfnamefont {David}\ \bibnamefont
  {Tom\'anek}}\ and\ \bibinfo {author} {\bibfnamefont {Andrii}\ \bibnamefont
  {Kyrylchuk}},\ }\bibfield  {title} {\enquote {\bibinfo {title} {Designing an
  all-carbon membrane for water desalination},}\ }\href {\doibase
  10.1103/PhysRevApplied.12.024054} {\bibfield  {journal} {\bibinfo  {journal}
  {Phys. Rev. Applied}\ }\textbf {\bibinfo {volume} {12}},\ \bibinfo {pages}
  {024054} (\bibinfo {year} {2019})}\BibitemShut {NoStop}%
\bibitem [{\citenamefont {Yang}\ \emph {et~al.}(2019)\citenamefont {Yang},
  \citenamefont {Yang}, \citenamefont {Liang}, \citenamefont {Gao},
  \citenamefont {Cheng}, \citenamefont {Li}, \citenamefont {Zou}, \citenamefont
  {Ma}, \citenamefont {Yuan},\ and\ \citenamefont {Duan}}]{Yang19}%
  \BibitemOpen
  \bibfield  {author} {\bibinfo {author} {\bibfnamefont {Yanbing}\ \bibnamefont
  {Yang}}, \bibinfo {author} {\bibfnamefont {Xiangdong}\ \bibnamefont {Yang}},
  \bibinfo {author} {\bibfnamefont {Ling}\ \bibnamefont {Liang}}, \bibinfo
  {author} {\bibfnamefont {Yuyan}\ \bibnamefont {Gao}}, \bibinfo {author}
  {\bibfnamefont {Huanyu}\ \bibnamefont {Cheng}}, \bibinfo {author}
  {\bibfnamefont {Xinming}\ \bibnamefont {Li}}, \bibinfo {author}
  {\bibfnamefont {Mingchu}\ \bibnamefont {Zou}}, \bibinfo {author}
  {\bibfnamefont {Renzhi}\ \bibnamefont {Ma}}, \bibinfo {author} {\bibfnamefont
  {Quan}\ \bibnamefont {Yuan}}, \ and\ \bibinfo {author} {\bibfnamefont
  {Xiangfeng}\ \bibnamefont {Duan}},\ }\bibfield  {title} {\enquote {\bibinfo
  {title} {Large-area graphene-nanomesh/carbon-nanotube hybrid membranes for
  ionic and molecular nanofiltration},}\ }\href {\doibase
  10.1126/science.aau5321} {\bibfield  {journal} {\bibinfo  {journal}
  {Science}\ }\textbf {\bibinfo {volume} {364}},\ \bibinfo {pages} {1057--1062}
  (\bibinfo {year} {2019})}\BibitemShut {NoStop}%
\bibitem [{\citenamefont {Rashid}\ and\ \citenamefont
  {Ralph}(2017)}]{Rashid2017}%
  \BibitemOpen
  \bibfield  {author} {\bibinfo {author} {\bibfnamefont {Md. Harun-Or}\
  \bibnamefont {Rashid}}\ and\ \bibinfo {author} {\bibfnamefont {Stephen~F.}\
  \bibnamefont {Ralph}},\ }\bibfield  {title} {\enquote {\bibinfo {title}
  {Carbon nanotube membranes: Synthesis, properties, and future filtration
  applications},}\ }\href {\doibase 10.3390/nano7050099} {\bibfield  {journal}
  {\bibinfo  {journal} {Nanomaterials}\ }\textbf {\bibinfo {volume} {7}},\
  \bibinfo {pages} {99} (\bibinfo {year} {2017})}\BibitemShut {NoStop}%
\bibitem [{\citenamefont {Chopra}\ \emph {et~al.}(1995)\citenamefont {Chopra},
  \citenamefont {Luyken}, \citenamefont {Cherrey}, \citenamefont {Crespi},
  \citenamefont {Cohen}, \citenamefont {Louie},\ and\ \citenamefont
  {Zettl}}]{Chopra1995}%
  \BibitemOpen
  \bibfield  {author} {\bibinfo {author} {\bibfnamefont {Nasreen~G.}\
  \bibnamefont {Chopra}}, \bibinfo {author} {\bibfnamefont {R.~J.}\
  \bibnamefont {Luyken}}, \bibinfo {author} {\bibfnamefont {K.}~\bibnamefont
  {Cherrey}}, \bibinfo {author} {\bibfnamefont {Vincent~H.}\ \bibnamefont
  {Crespi}}, \bibinfo {author} {\bibfnamefont {Marvin~L.}\ \bibnamefont
  {Cohen}}, \bibinfo {author} {\bibfnamefont {Steven~G.}\ \bibnamefont
  {Louie}}, \ and\ \bibinfo {author} {\bibfnamefont {A.}~\bibnamefont
  {Zettl}},\ }\bibfield  {title} {\enquote {\bibinfo {title} {Boron nitride
  nanotubes},}\ }\href {\doibase 10.1126/science.269.5226.966} {\bibfield
  {journal} {\bibinfo  {journal} {Science}\ }\textbf {\bibinfo {volume}
  {269}},\ \bibinfo {pages} {966--967} (\bibinfo {year} {1995})}\BibitemShut
  {NoStop}%
\bibitem [{\citenamefont {Faucher}\ \emph {et~al.}(2019)\citenamefont
  {Faucher}, \citenamefont {Aluru}, \citenamefont {Bazant}, \citenamefont
  {Blankschtein}, \citenamefont {Brozena}, \citenamefont {Cumings},
  \citenamefont {{Pedro de Souza}}, \citenamefont {Elimelech}, \citenamefont
  {Epsztein}, \citenamefont {Fourkas}, \citenamefont {Rajan}, \citenamefont
  {Kulik}, \citenamefont {Levy}, \citenamefont {Majumdar}, \citenamefont
  {Martin}, \citenamefont {McEldrew}, \citenamefont {Misra}, \citenamefont
  {Noy}, \citenamefont {Pham}, \citenamefont {Reed}, \citenamefont {Schwegler},
  \citenamefont {Siwy}, \citenamefont {Wang},\ and\ \citenamefont
  {Strano}}]{Faucher2019}%
  \BibitemOpen
  \bibfield  {author} {\bibinfo {author} {\bibfnamefont {Samuel}\ \bibnamefont
  {Faucher}}, \bibinfo {author} {\bibfnamefont {Narayana}\ \bibnamefont
  {Aluru}}, \bibinfo {author} {\bibfnamefont {Martin~Z.}\ \bibnamefont
  {Bazant}}, \bibinfo {author} {\bibfnamefont {Daniel}\ \bibnamefont
  {Blankschtein}}, \bibinfo {author} {\bibfnamefont {Alexandra~H.}\
  \bibnamefont {Brozena}}, \bibinfo {author} {\bibfnamefont {John}\
  \bibnamefont {Cumings}}, \bibinfo {author} {\bibfnamefont {J.}~\bibnamefont
  {{Pedro de Souza}}}, \bibinfo {author} {\bibfnamefont {Menachem}\
  \bibnamefont {Elimelech}}, \bibinfo {author} {\bibfnamefont {Razi}\
  \bibnamefont {Epsztein}}, \bibinfo {author} {\bibfnamefont {John~T.}\
  \bibnamefont {Fourkas}}, \bibinfo {author} {\bibfnamefont {Ananth~Govind}\
  \bibnamefont {Rajan}}, \bibinfo {author} {\bibfnamefont {Heather~J.}\
  \bibnamefont {Kulik}}, \bibinfo {author} {\bibfnamefont {Amir}\ \bibnamefont
  {Levy}}, \bibinfo {author} {\bibfnamefont {Arun}\ \bibnamefont {Majumdar}},
  \bibinfo {author} {\bibfnamefont {Charles}\ \bibnamefont {Martin}}, \bibinfo
  {author} {\bibfnamefont {Michael}\ \bibnamefont {McEldrew}}, \bibinfo
  {author} {\bibfnamefont {Rahul~Prasanna}\ \bibnamefont {Misra}}, \bibinfo
  {author} {\bibfnamefont {Aleksandr}\ \bibnamefont {Noy}}, \bibinfo {author}
  {\bibfnamefont {Tuan~Anh}\ \bibnamefont {Pham}}, \bibinfo {author}
  {\bibfnamefont {Mark}\ \bibnamefont {Reed}}, \bibinfo {author} {\bibfnamefont
  {Eric}\ \bibnamefont {Schwegler}}, \bibinfo {author} {\bibfnamefont
  {Zuzanna}\ \bibnamefont {Siwy}}, \bibinfo {author} {\bibfnamefont {Yuhuang}\
  \bibnamefont {Wang}}, \ and\ \bibinfo {author} {\bibfnamefont {Michael}\
  \bibnamefont {Strano}},\ }\bibfield  {title} {\enquote {\bibinfo {title}
  {Critical knowledge gaps in mass transport through single-digit nanopores: A
  review and perspective},}\ }\href {\doibase 10.1021/acs.jpcc.9b02178}
  {\bibfield  {journal} {\bibinfo  {journal} {J. Phys. Chem. C}\ }\textbf
  {\bibinfo {volume} {123}},\ \bibinfo {pages} {21309--21326} (\bibinfo {year}
  {2019})}\BibitemShut {NoStop}%
\bibitem [{\citenamefont {Kalra}\ \emph {et~al.}(2003)\citenamefont {Kalra},
  \citenamefont {Garde},\ and\ \citenamefont {Hummer}}]{Kalra2003}%
  \BibitemOpen
  \bibfield  {author} {\bibinfo {author} {\bibfnamefont {Amrit}\ \bibnamefont
  {Kalra}}, \bibinfo {author} {\bibfnamefont {Shekhar}\ \bibnamefont {Garde}},
  \ and\ \bibinfo {author} {\bibfnamefont {Gerhard}\ \bibnamefont {Hummer}},\
  }\bibfield  {title} {\enquote {\bibinfo {title} {(cul-id:1887995) from the
  cover: Osmotic water transport through carbon nanotube membranes},}\ }\href
  {\doibase 10.1073/pnas.1633354100} {\bibfield  {journal} {\bibinfo  {journal}
  {Proc. Natl. Acad. Sci. U.S.A.}\ }\textbf {\bibinfo {volume} {100}},\
  \bibinfo {pages} {10175--10180} (\bibinfo {year} {2003})}\BibitemShut
  {NoStop}%
\bibitem [{\citenamefont {Zhu}\ and\ \citenamefont {Schulten}(2003)}]{Zhu2003}%
  \BibitemOpen
  \bibfield  {author} {\bibinfo {author} {\bibfnamefont {Fangqiang}\
  \bibnamefont {Zhu}}\ and\ \bibinfo {author} {\bibfnamefont {Klaus}\
  \bibnamefont {Schulten}},\ }\bibfield  {title} {\enquote {\bibinfo {title}
  {Water and proton conduction through carbon nanotubes as models for
  biological channels},}\ }\href {\doibase 10.1016/S0006-3495(03)74469-5}
  {\bibfield  {journal} {\bibinfo  {journal} {Biophys. J.}\ }\textbf {\bibinfo
  {volume} {85}},\ \bibinfo {pages} {236--244} (\bibinfo {year}
  {2003})}\BibitemShut {NoStop}%
\bibitem [{\citenamefont {Thomas}\ and\ \citenamefont
  {McGaughey}(2009)}]{Thomas2009}%
  \BibitemOpen
  \bibfield  {author} {\bibinfo {author} {\bibfnamefont {John~A.}\ \bibnamefont
  {Thomas}}\ and\ \bibinfo {author} {\bibfnamefont {Alan J.~H.}\ \bibnamefont
  {McGaughey}},\ }\bibfield  {title} {\enquote {\bibinfo {title} {Water flow in
  carbon nanotubes: Transition to subcontinuum transport},}\ }\href {\doibase
  10.1103/PhysRevLett.102.184502} {\bibfield  {journal} {\bibinfo  {journal}
  {Phys. Rev. Lett.}\ }\textbf {\bibinfo {volume} {102}},\ \bibinfo {pages}
  {184502} (\bibinfo {year} {2009})}\BibitemShut {NoStop}%
\bibitem [{\citenamefont {Hummer}\ \emph {et~al.}(2001)\citenamefont {Hummer},
  \citenamefont {Rasaiah},\ and\ \citenamefont {Noworyta}}]{Hummer2001}%
  \BibitemOpen
  \bibfield  {author} {\bibinfo {author} {\bibfnamefont {G.}~\bibnamefont
  {Hummer}}, \bibinfo {author} {\bibfnamefont {J.~C.}\ \bibnamefont {Rasaiah}},
  \ and\ \bibinfo {author} {\bibfnamefont {J.~P.}\ \bibnamefont {Noworyta}},\
  }\bibfield  {title} {\enquote {\bibinfo {title} {Water conduction through the
  hydrophobic channel of a carbon nanotube},}\ }\href {\doibase
  10.1038/35102535} {\bibfield  {journal} {\bibinfo  {journal} {Nature}\
  }\textbf {\bibinfo {volume} {414}},\ \bibinfo {pages} {188--190} (\bibinfo
  {year} {2001})}\BibitemShut {NoStop}%
\bibitem [{\citenamefont {Misra}\ and\ \citenamefont
  {Blankschtein}(2017)}]{Misra2017a}%
  \BibitemOpen
  \bibfield  {author} {\bibinfo {author} {\bibfnamefont {Rahul~Prasanna}\
  \bibnamefont {Misra}}\ and\ \bibinfo {author} {\bibfnamefont {Daniel}\
  \bibnamefont {Blankschtein}},\ }\bibfield  {title} {\enquote {\bibinfo
  {title} {Insights on the role of many-body polarization effects in the
  wetting of graphitic surfaces by water},}\ }\href {\doibase
  10.1021/acs.jpcc.7b08891} {\bibfield  {journal} {\bibinfo  {journal} {J.
  Phys. Chem. C}\ }\textbf {\bibinfo {volume} {121}},\ \bibinfo {pages}
  {28166--28179} (\bibinfo {year} {2017})}\BibitemShut {NoStop}%
\bibitem [{\citenamefont {Mann}\ and\ \citenamefont {Halls}(2003)}]{Mann2003}%
  \BibitemOpen
  \bibfield  {author} {\bibinfo {author} {\bibfnamefont {David~J.}\
  \bibnamefont {Mann}}\ and\ \bibinfo {author} {\bibfnamefont {Mathew~D.}\
  \bibnamefont {Halls}},\ }\bibfield  {title} {\enquote {\bibinfo {title}
  {Water alignment and proton conduction inside carbon nanotubes},}\ }\href
  {\doibase 10.1103/PhysRevLett.90.195503} {\bibfield  {journal} {\bibinfo
  {journal} {Phys. Rev. Lett.}\ }\textbf {\bibinfo {volume} {90}},\ \bibinfo
  {pages} {195503} (\bibinfo {year} {2003})}\BibitemShut {NoStop}%
\bibitem [{\citenamefont {Dellago}\ \emph {et~al.}(2003)\citenamefont
  {Dellago}, \citenamefont {Naor},\ and\ \citenamefont {Hummer}}]{Dellago2003}%
  \BibitemOpen
  \bibfield  {author} {\bibinfo {author} {\bibfnamefont {Christoph}\
  \bibnamefont {Dellago}}, \bibinfo {author} {\bibfnamefont {Mor~M.}\
  \bibnamefont {Naor}}, \ and\ \bibinfo {author} {\bibfnamefont {Gerhard}\
  \bibnamefont {Hummer}},\ }\bibfield  {title} {\enquote {\bibinfo {title}
  {Proton transport through water-filled carbon nanotubes},}\ }\href {\doibase
  10.1103/PhysRevLett.90.105902} {\bibfield  {journal} {\bibinfo  {journal}
  {Phys. Rev. Lett.}\ }\textbf {\bibinfo {volume} {90}},\ \bibinfo {pages}
  {105902} (\bibinfo {year} {2003})}\BibitemShut {NoStop}%
\bibitem [{\citenamefont {Cambr\'e}\ \emph {et~al.}(2010)\citenamefont
  {Cambr\'e}, \citenamefont {Schoeters}, \citenamefont {Luyckx}, \citenamefont
  {Goovaerts},\ and\ \citenamefont {Wenseleers}}]{Cambre2010}%
  \BibitemOpen
  \bibfield  {author} {\bibinfo {author} {\bibfnamefont {Sofie}\ \bibnamefont
  {Cambr\'e}}, \bibinfo {author} {\bibfnamefont {Bob}\ \bibnamefont
  {Schoeters}}, \bibinfo {author} {\bibfnamefont {Sten}\ \bibnamefont
  {Luyckx}}, \bibinfo {author} {\bibfnamefont {Etienne}\ \bibnamefont
  {Goovaerts}}, \ and\ \bibinfo {author} {\bibfnamefont {Wim}\ \bibnamefont
  {Wenseleers}},\ }\bibfield  {title} {\enquote {\bibinfo {title} {Experimental
  observation of single-file water filling of thin single-wall carbon nanotubes
  down to chiral index (5,3)},}\ }\href {\doibase
  10.1103/PhysRevLett.104.207401} {\bibfield  {journal} {\bibinfo  {journal}
  {Phys. Rev. Lett.}\ }\textbf {\bibinfo {volume} {104}},\ \bibinfo {pages}
  {207401} (\bibinfo {year} {2010})}\BibitemShut {NoStop}%
\bibitem [{\citenamefont {Kolesnikov}\ \emph {et~al.}(2004)\citenamefont
  {Kolesnikov}, \citenamefont {Zanotti}, \citenamefont {Loong}, \citenamefont
  {Thiyagarajan}, \citenamefont {Moravsky}, \citenamefont {Loutfy},\ and\
  \citenamefont {Burnham}}]{Kolesnikov2004}%
  \BibitemOpen
  \bibfield  {author} {\bibinfo {author} {\bibfnamefont {Alexander~I.}\
  \bibnamefont {Kolesnikov}}, \bibinfo {author} {\bibfnamefont {Jean~Marc}\
  \bibnamefont {Zanotti}}, \bibinfo {author} {\bibfnamefont {Chun~Keung}\
  \bibnamefont {Loong}}, \bibinfo {author} {\bibfnamefont {Pappannan}\
  \bibnamefont {Thiyagarajan}}, \bibinfo {author} {\bibfnamefont
  {Alexander~P.}\ \bibnamefont {Moravsky}}, \bibinfo {author} {\bibfnamefont
  {Raouf~O.}\ \bibnamefont {Loutfy}}, \ and\ \bibinfo {author} {\bibfnamefont
  {Christian~J.}\ \bibnamefont {Burnham}},\ }\bibfield  {title} {\enquote
  {\bibinfo {title} {Anomalously soft dynamics of water in a nanotube: A
  revelation of nanoscale confinement},}\ }\href {\doibase
  10.1103/PhysRevLett.93.035503} {\bibfield  {journal} {\bibinfo  {journal}
  {Phys. Rev. Lett.}\ }\textbf {\bibinfo {volume} {93}},\ \bibinfo {pages}
  {035503--1} (\bibinfo {year} {2004})}\BibitemShut {NoStop}%
\bibitem [{\citenamefont {Liu}\ \emph {et~al.}(2005)\citenamefont {Liu},
  \citenamefont {Wang}, \citenamefont {Wu},\ and\ \citenamefont
  {Zhang}}]{Liu2005}%
  \BibitemOpen
  \bibfield  {author} {\bibinfo {author} {\bibfnamefont {Yingchun}\
  \bibnamefont {Liu}}, \bibinfo {author} {\bibfnamefont {Qi}~\bibnamefont
  {Wang}}, \bibinfo {author} {\bibfnamefont {Tao}\ \bibnamefont {Wu}}, \ and\
  \bibinfo {author} {\bibfnamefont {Li}~\bibnamefont {Zhang}},\ }\bibfield
  {title} {\enquote {\bibinfo {title} {Fluid structure and transport properties
  of water inside carbon nanotubes},}\ }\href {\doibase 10.1063/1.2131070}
  {\bibfield  {journal} {\bibinfo  {journal} {J. Chem. Phys.}\ }\textbf
  {\bibinfo {volume} {123}},\ \bibinfo {pages} {234701} (\bibinfo {year}
  {2005})}\BibitemShut {NoStop}%
\bibitem [{\citenamefont {Koga}\ \emph {et~al.}(2001)\citenamefont {Koga},
  \citenamefont {Gao}, \citenamefont {Tanaka},\ and\ \citenamefont
  {Zeng}}]{Koga2001}%
  \BibitemOpen
  \bibfield  {author} {\bibinfo {author} {\bibfnamefont {Kenichiro}\
  \bibnamefont {Koga}}, \bibinfo {author} {\bibfnamefont {G.~T.}\ \bibnamefont
  {Gao}}, \bibinfo {author} {\bibfnamefont {Hideki}\ \bibnamefont {Tanaka}}, \
  and\ \bibinfo {author} {\bibfnamefont {X.~C.}\ \bibnamefont {Zeng}},\
  }\bibfield  {title} {\enquote {\bibinfo {title} {Formation of ordered ice
  nanotubes inside carbon nanotubes},}\ }\href {\doibase 10.1038/35090532}
  {\bibfield  {journal} {\bibinfo  {journal} {Nature}\ }\textbf {\bibinfo
  {volume} {412}},\ \bibinfo {pages} {802--805} (\bibinfo {year}
  {2001})}\BibitemShut {NoStop}%
\bibitem [{\citenamefont {Cicero}\ \emph {et~al.}(2008)\citenamefont {Cicero},
  \citenamefont {Grossman}, \citenamefont {Schwegler}, \citenamefont {Gygi},\
  and\ \citenamefont {Galli}}]{Cicero2008}%
  \BibitemOpen
  \bibfield  {author} {\bibinfo {author} {\bibfnamefont {Giancarlo}\
  \bibnamefont {Cicero}}, \bibinfo {author} {\bibfnamefont {Jeffrey~C.}\
  \bibnamefont {Grossman}}, \bibinfo {author} {\bibfnamefont {Eric}\
  \bibnamefont {Schwegler}}, \bibinfo {author} {\bibfnamefont {Francois}\
  \bibnamefont {Gygi}}, \ and\ \bibinfo {author} {\bibfnamefont {Giulia}\
  \bibnamefont {Galli}},\ }\bibfield  {title} {\enquote {\bibinfo {title}
  {Water confined in nanotubes and between graphene sheets: a first principle
  study.}}\ }\href {\doibase 10.1021/ja074418+} {\bibfield  {journal} {\bibinfo
   {journal} {J. Am. Chem. Soc.}\ }\textbf {\bibinfo {volume} {130}},\ \bibinfo
  {pages} {1871--8} (\bibinfo {year} {2008})}\BibitemShut {NoStop}%
\bibitem [{\citenamefont {Byl}\ \emph {et~al.}(2006)\citenamefont {Byl},
  \citenamefont {Liu}, \citenamefont {Wang}, \citenamefont {Yim}, \citenamefont
  {Johnson},\ and\ \citenamefont {Yates}}]{Byl2006}%
  \BibitemOpen
  \bibfield  {author} {\bibinfo {author} {\bibfnamefont {Oleg}\ \bibnamefont
  {Byl}}, \bibinfo {author} {\bibfnamefont {Jin~Chen}\ \bibnamefont {Liu}},
  \bibinfo {author} {\bibfnamefont {Yang}\ \bibnamefont {Wang}}, \bibinfo
  {author} {\bibfnamefont {Wai~Leung}\ \bibnamefont {Yim}}, \bibinfo {author}
  {\bibfnamefont {J.~Karl}\ \bibnamefont {Johnson}}, \ and\ \bibinfo {author}
  {\bibfnamefont {John~T.}\ \bibnamefont {Yates}},\ }\bibfield  {title}
  {\enquote {\bibinfo {title} {Unusual hydrogen bonding in water-filled carbon
  nanotubes},}\ }\href {\doibase 10.1021/ja057856u} {\bibfield  {journal}
  {\bibinfo  {journal} {J. Am. Chem. Soc.}\ }\textbf {\bibinfo {volume}
  {128}},\ \bibinfo {pages} {12090--12097} (\bibinfo {year}
  {2006})}\BibitemShut {NoStop}%
\bibitem [{\citenamefont {Maniwa}\ \emph {et~al.}(2005)\citenamefont {Maniwa},
  \citenamefont {Kataura}, \citenamefont {Abe}, \citenamefont {Udaka},
  \citenamefont {Suzuki}, \citenamefont {Achiba}, \citenamefont {Kira},
  \citenamefont {Matsuda}, \citenamefont {Kadowaki},\ and\ \citenamefont
  {Okabe}}]{Maniwa2005}%
  \BibitemOpen
  \bibfield  {author} {\bibinfo {author} {\bibfnamefont {Yutaka}\ \bibnamefont
  {Maniwa}}, \bibinfo {author} {\bibfnamefont {Hiromichi}\ \bibnamefont
  {Kataura}}, \bibinfo {author} {\bibfnamefont {Masatoshi}\ \bibnamefont
  {Abe}}, \bibinfo {author} {\bibfnamefont {Akiko}\ \bibnamefont {Udaka}},
  \bibinfo {author} {\bibfnamefont {Shinzo}\ \bibnamefont {Suzuki}}, \bibinfo
  {author} {\bibfnamefont {Yohji}\ \bibnamefont {Achiba}}, \bibinfo {author}
  {\bibfnamefont {Hiroshi}\ \bibnamefont {Kira}}, \bibinfo {author}
  {\bibfnamefont {Kazuyuki}\ \bibnamefont {Matsuda}}, \bibinfo {author}
  {\bibfnamefont {Hiroaki}\ \bibnamefont {Kadowaki}}, \ and\ \bibinfo {author}
  {\bibfnamefont {Yutaka}\ \bibnamefont {Okabe}},\ }\bibfield  {title}
  {\enquote {\bibinfo {title} {Ordered water inside carbon nanotubes: Formation
  of pentagonal to octagonal ice-nanotubes},}\ }\href {\doibase
  10.1016/j.cplett.2004.11.112} {\bibfield  {journal} {\bibinfo  {journal}
  {Chem. Phys. Lett.}\ }\textbf {\bibinfo {volume} {401}},\ \bibinfo {pages}
  {534--538} (\bibinfo {year} {2005})}\BibitemShut {NoStop}%
\bibitem [{\citenamefont {Holt}(2006)}]{Holt2006a}%
  \BibitemOpen
  \bibfield  {author} {\bibinfo {author} {\bibfnamefont {J.~K.}\ \bibnamefont
  {Holt}},\ }\bibfield  {title} {\enquote {\bibinfo {title} {Fast mass
  transport through sub-2-nanometer carbon nanotubes},}\ }\href {\doibase
  10.1126/science.1126298} {\bibfield  {journal} {\bibinfo  {journal}
  {Science}\ }\textbf {\bibinfo {volume} {312}},\ \bibinfo {pages} {1034--1037}
  (\bibinfo {year} {2006})}\BibitemShut {NoStop}%
\bibitem [{\citenamefont {Eijkel}(2007)}]{Eijkel2007}%
  \BibitemOpen
  \bibfield  {author} {\bibinfo {author} {\bibfnamefont {J.}~\bibnamefont
  {Eijkel}},\ }\bibfield  {title} {\enquote {\bibinfo {title} {Liquid slip in
  micro- and nanofluidics: Recent research and its possible implications},}\
  }\href {\doibase 10.1039/b700364c} {\bibfield  {journal} {\bibinfo  {journal}
  {Lab. Chip.}\ }\textbf {\bibinfo {volume} {7}},\ \bibinfo {pages} {299--301}
  (\bibinfo {year} {2007})}\BibitemShut {NoStop}%
\bibitem [{\citenamefont {Lauga}\ \emph {et~al.}(2007)\citenamefont {Lauga},
  \citenamefont {Brenner},\ and\ \citenamefont {Stone}}]{Lauga2007}%
  \BibitemOpen
  \bibfield  {author} {\bibinfo {author} {\bibfnamefont {Eric}\ \bibnamefont
  {Lauga}}, \bibinfo {author} {\bibfnamefont {Michael}\ \bibnamefont
  {Brenner}}, \ and\ \bibinfo {author} {\bibfnamefont {Howard}\ \bibnamefont
  {Stone}},\ }\bibfield  {title} {\enquote {\bibinfo {title} {Microfluidics:
  The no-slip boundary condition},}\ }in\ \href {\doibase
  10.1007/978-3-540-30299-5_19} {\emph {\bibinfo {booktitle} {Springer Handbook
  of Experimental Fluid Mechanics}}}\ (\bibinfo  {publisher} {Springer Berlin
  Heidelberg},\ \bibinfo {address} {Berlin, Heidelberg},\ \bibinfo {year}
  {2007})\ pp.\ \bibinfo {pages} {1219--1240}\BibitemShut {NoStop}%
\bibitem [{\citenamefont {Whitby}\ \emph {et~al.}(2008)\citenamefont {Whitby},
  \citenamefont {Cagnon}, \citenamefont {Thanou},\ and\ \citenamefont
  {Quirke}}]{Whitby2008}%
  \BibitemOpen
  \bibfield  {author} {\bibinfo {author} {\bibfnamefont {Max}\ \bibnamefont
  {Whitby}}, \bibinfo {author} {\bibfnamefont {Laurent}\ \bibnamefont
  {Cagnon}}, \bibinfo {author} {\bibfnamefont {Maya}\ \bibnamefont {Thanou}}, \
  and\ \bibinfo {author} {\bibfnamefont {Nick}\ \bibnamefont {Quirke}},\
  }\bibfield  {title} {\enquote {\bibinfo {title} {Enhanced fluid flow through
  nanoscale carbon pipes},}\ }\href {\doibase 10.1021/nl080705f} {\bibfield
  {journal} {\bibinfo  {journal} {Nano Lett.}\ }\textbf {\bibinfo {volume}
  {8}},\ \bibinfo {pages} {2632--2637} (\bibinfo {year} {2008})}\BibitemShut
  {NoStop}%
\bibitem [{\citenamefont {Tsimpanogiannis}\ \emph {et~al.}(2019)\citenamefont
  {Tsimpanogiannis}, \citenamefont {Moultos}, \citenamefont {Franco},
  \citenamefont {Spera}, \citenamefont {Erd{\H{o}}s},\ and\ \citenamefont
  {Economou}}]{Tsimpanogiannis2019}%
  \BibitemOpen
  \bibfield  {author} {\bibinfo {author} {\bibfnamefont {Ioannis~N.}\
  \bibnamefont {Tsimpanogiannis}}, \bibinfo {author} {\bibfnamefont
  {Othonas~A.}\ \bibnamefont {Moultos}}, \bibinfo {author} {\bibfnamefont
  {Lu{\'{i}}s~F.M.}\ \bibnamefont {Franco}}, \bibinfo {author} {\bibfnamefont
  {Marcelle B.de~M.}\ \bibnamefont {Spera}}, \bibinfo {author} {\bibfnamefont
  {M{\'{a}}t{\'{e}}}\ \bibnamefont {Erd{\H{o}}s}}, \ and\ \bibinfo {author}
  {\bibfnamefont {Ioannis~G.}\ \bibnamefont {Economou}},\ }\bibfield  {title}
  {\enquote {\bibinfo {title} {Self-diffusion coefficient of bulk and confined
  water: a critical review of classical molecular simulation studies},}\ }\href
  {\doibase 10.1080/08927022.2018.1511903} {\bibfield  {journal} {\bibinfo
  {journal} {Mol. Simul.}\ }\textbf {\bibinfo {volume} {45}},\ \bibinfo {pages}
  {425--453} (\bibinfo {year} {2019})}\BibitemShut {NoStop}%
\bibitem [{\citenamefont {Quandt}\ \emph {et~al.}()\citenamefont {Quandt},
  \citenamefont {Kyrylchuk}, \citenamefont {Seifert},\ and\ \citenamefont
  {Tom\'anek}}]{flow20}%
  \BibitemOpen
  \bibfield  {author} {\bibinfo {author} {\bibfnamefont {Alexander}\
  \bibnamefont {Quandt}}, \bibinfo {author} {\bibfnamefont {Andrii}\
  \bibnamefont {Kyrylchuk}}, \bibinfo {author} {\bibfnamefont {Gotthard}\
  \bibnamefont {Seifert}}, \ and\ \bibinfo {author} {\bibfnamefont {David}\
  \bibnamefont {Tom\'anek}},\ }\href@noop {} {\enquote {\bibinfo {title} {Water
  flow through a graphite oxide membrane: A phenomenological description},}\
  }\bibinfo {note} {(Unpublished)}\BibitemShut {NoStop}%
\bibitem [{\citenamefont {Artacho}\ \emph {et~al.}(2008)\citenamefont
  {Artacho}, \citenamefont {Anglada}, \citenamefont {Dieguez}, \citenamefont
  {Gale}, \citenamefont {Garcia}, \citenamefont {Junquera}, \citenamefont
  {Martin}, \citenamefont {Ordejon}, \citenamefont {Pruneda}, \citenamefont
  {Sanchez-Portal},\ and\ \citenamefont {Soler}}]{SIESTA}%
  \BibitemOpen
  \bibfield  {author} {\bibinfo {author} {\bibfnamefont {Emilio}\ \bibnamefont
  {Artacho}}, \bibinfo {author} {\bibfnamefont {E.}~\bibnamefont {Anglada}},
  \bibinfo {author} {\bibfnamefont {O.}~\bibnamefont {Dieguez}}, \bibinfo
  {author} {\bibfnamefont {J.~D.}\ \bibnamefont {Gale}}, \bibinfo {author}
  {\bibfnamefont {A.}~\bibnamefont {Garcia}}, \bibinfo {author} {\bibfnamefont
  {J.}~\bibnamefont {Junquera}}, \bibinfo {author} {\bibfnamefont {R.~M.}\
  \bibnamefont {Martin}}, \bibinfo {author} {\bibfnamefont {P.}~\bibnamefont
  {Ordejon}}, \bibinfo {author} {\bibfnamefont {J.~M.}\ \bibnamefont
  {Pruneda}}, \bibinfo {author} {\bibfnamefont {D.}~\bibnamefont
  {Sanchez-Portal}}, \ and\ \bibinfo {author} {\bibfnamefont {J.~M.}\
  \bibnamefont {Soler}},\ }\bibfield  {title} {\enquote {\bibinfo {title} {The
  siesta method; developments and applicability},}\ }\href {\doibase
  10.1088/0953-8984/20/6/064208} {\bibfield  {journal} {\bibinfo  {journal} {J.
  Phys. Cond. Mat.}\ }\textbf {\bibinfo {volume} {20}},\ \bibinfo {pages}
  {064208} (\bibinfo {year} {2008})}\BibitemShut {NoStop}%
\bibitem [{\citenamefont {Perdew}\ \emph {et~al.}(1996)\citenamefont {Perdew},
  \citenamefont {Burke},\ and\ \citenamefont {Ernzerhof}}]{PBE}%
  \BibitemOpen
  \bibfield  {author} {\bibinfo {author} {\bibfnamefont {John~P.}\ \bibnamefont
  {Perdew}}, \bibinfo {author} {\bibfnamefont {Kieron}\ \bibnamefont {Burke}},
  \ and\ \bibinfo {author} {\bibfnamefont {Matthias}\ \bibnamefont
  {Ernzerhof}},\ }\bibfield  {title} {\enquote {\bibinfo {title} {Generalized
  gradient approximation made simple},}\ }\href {\doibase
  10.1103/PhysRevLett.77.3865} {\bibfield  {journal} {\bibinfo  {journal}
  {Phys. Rev. Lett.}\ }\textbf {\bibinfo {volume} {77}},\ \bibinfo {pages}
  {3865--3868} (\bibinfo {year} {1996})}\BibitemShut {NoStop}%
\bibitem [{\citenamefont {Troullier}\ and\ \citenamefont
  {Martins}(1991)}]{Troullier91}%
  \BibitemOpen
  \bibfield  {author} {\bibinfo {author} {\bibfnamefont {N.}~\bibnamefont
  {Troullier}}\ and\ \bibinfo {author} {\bibfnamefont {Jos\'e~Lu\'{\i}s}\
  \bibnamefont {Martins}},\ }\bibfield  {title} {\enquote {\bibinfo {title}
  {Efficient pseudopotentials for plane-wave calculations},}\ }\href {\doibase
  10.1103/PhysRevB.43.1993} {\bibfield  {journal} {\bibinfo  {journal} {Phys.
  Rev. B}\ }\textbf {\bibinfo {volume} {43}},\ \bibinfo {pages} {1993--2006}
  (\bibinfo {year} {1991})}\BibitemShut {NoStop}%
\bibitem [{\citenamefont {Monkhorst}\ and\ \citenamefont
  {Pack}(1976)}]{Monkhorst-Pack76}%
  \BibitemOpen
  \bibfield  {author} {\bibinfo {author} {\bibfnamefont {Hendrik~J.}\
  \bibnamefont {Monkhorst}}\ and\ \bibinfo {author} {\bibfnamefont {James~D.}\
  \bibnamefont {Pack}},\ }\bibfield  {title} {\enquote {\bibinfo {title}
  {Special points for brillouin-zone integrations},}\ }\href {\doibase
  10.1103/PhysRevB.13.5188} {\bibfield  {journal} {\bibinfo  {journal} {Phys.
  Rev. B}\ }\textbf {\bibinfo {volume} {13}},\ \bibinfo {pages} {5188--5192}
  (\bibinfo {year} {1976})}\BibitemShut {NoStop}%
\bibitem [{\citenamefont {Hestenes}\ and\ \citenamefont
  {Stiefel}(1952)}]{Hestenes1952}%
  \BibitemOpen
  \bibfield  {author} {\bibinfo {author} {\bibfnamefont {M.R.}\ \bibnamefont
  {Hestenes}}\ and\ \bibinfo {author} {\bibfnamefont {E.}~\bibnamefont
  {Stiefel}},\ }\bibfield  {title} {\enquote {\bibinfo {title} {Methods of
  conjugate gradients for solving linear systems},}\ }\href {\doibase
  10.6028/jres.049.044} {\bibfield  {journal} {\bibinfo  {journal} {J. Res.
  Natl. Bur. Stand. (U. S.)}\ }\textbf {\bibinfo {volume} {49}},\ \bibinfo
  {pages} {409} (\bibinfo {year} {1952})}\BibitemShut {NoStop}%
\bibitem [{\citenamefont {Ambrosetti}\ and\ \citenamefont
  {Silvestrelli}(2011)}]{Ambrosetti2011}%
  \BibitemOpen
  \bibfield  {author} {\bibinfo {author} {\bibfnamefont {A.}~\bibnamefont
  {Ambrosetti}}\ and\ \bibinfo {author} {\bibfnamefont {P.~L.}\ \bibnamefont
  {Silvestrelli}},\ }\bibfield  {title} {\enquote {\bibinfo {title} {Adsorption
  of rare-gas atoms and water on graphite and graphene by van der
  waals-corrected density functional theory},}\ }\href {\doibase
  10.1021/jp110669p} {\bibfield  {journal} {\bibinfo  {journal} {J. Phys. Chem.
  C}\ }\textbf {\bibinfo {volume} {115}},\ \bibinfo {pages} {3695--3702}
  (\bibinfo {year} {2011})}\BibitemShut {NoStop}%
\bibitem [{\citenamefont {Kannam}\ \emph {et~al.}(2013)\citenamefont {Kannam},
  \citenamefont {Todd}, \citenamefont {Hansen},\ and\ \citenamefont
  {Daivis}}]{Kannam2013}%
  \BibitemOpen
  \bibfield  {author} {\bibinfo {author} {\bibfnamefont {Sridhar~Kumar}\
  \bibnamefont {Kannam}}, \bibinfo {author} {\bibfnamefont {B.~D.}\
  \bibnamefont {Todd}}, \bibinfo {author} {\bibfnamefont {J.~S.}\ \bibnamefont
  {Hansen}}, \ and\ \bibinfo {author} {\bibfnamefont {Peter~J.}\ \bibnamefont
  {Daivis}},\ }\bibfield  {title} {\enquote {\bibinfo {title} {How fast does
  water flow in carbon nanotubes?}}\ }\href {\doibase 10.1063/1.4793396}
  {\bibfield  {journal} {\bibinfo  {journal} {J. Chem. Phys.}\ }\textbf
  {\bibinfo {volume} {138}},\ \bibinfo {pages} {094701} (\bibinfo {year}
  {2013})}\BibitemShut {NoStop}%
\bibitem [{\citenamefont {Falk}\ \emph {et~al.}(2010)\citenamefont {Falk},
  \citenamefont {Sedlmeier}, \citenamefont {Joly}, \citenamefont {Netz},\ and\
  \citenamefont {Bocquet}}]{Falk2010}%
  \BibitemOpen
  \bibfield  {author} {\bibinfo {author} {\bibfnamefont {Kerstin}\ \bibnamefont
  {Falk}}, \bibinfo {author} {\bibfnamefont {Felix}\ \bibnamefont {Sedlmeier}},
  \bibinfo {author} {\bibfnamefont {Laurent}\ \bibnamefont {Joly}}, \bibinfo
  {author} {\bibfnamefont {Roland~R.}\ \bibnamefont {Netz}}, \ and\ \bibinfo
  {author} {\bibfnamefont {Lyd\'{e}ric}\ \bibnamefont {Bocquet}},\ }\bibfield
  {title} {\enquote {\bibinfo {title} {Molecular origin of fast water transport
  in carbon nanotube membranes: Superlubricity versus curvature dependent
  friction},}\ }\href {\doibase 10.1021/nl1021046} {\bibfield  {journal}
  {\bibinfo  {journal} {Nano Lett.}\ }\textbf {\bibinfo {volume} {10}},\
  \bibinfo {pages} {4067--4073} (\bibinfo {year} {2010})}\BibitemShut {NoStop}%
\bibitem [{\citenamefont {Boukhvalov}\ \emph {et~al.}(2013)\citenamefont
  {Boukhvalov}, \citenamefont {Katsnelson},\ and\ \citenamefont
  {Son}}]{Boukhvalov2013}%
  \BibitemOpen
  \bibfield  {author} {\bibinfo {author} {\bibfnamefont {Danil~W.}\
  \bibnamefont {Boukhvalov}}, \bibinfo {author} {\bibfnamefont {Mikhail~I.}\
  \bibnamefont {Katsnelson}}, \ and\ \bibinfo {author} {\bibfnamefont
  {Young~Woo}\ \bibnamefont {Son}},\ }\bibfield  {title} {\enquote {\bibinfo
  {title} {Origin of anomalous water permeation through graphene oxide
  membrane},}\ }\href {\doibase 10.1021/nl4020292} {\bibfield  {journal}
  {\bibinfo  {journal} {Nano Lett.}\ }\textbf {\bibinfo {volume} {13}},\
  \bibinfo {pages} {3930--3935} (\bibinfo {year} {2013})}\BibitemShut {NoStop}%
\bibitem [{\citenamefont {Wei}\ and\ \citenamefont {Luo}(2018)}]{Wei2018}%
  \BibitemOpen
  \bibfield  {author} {\bibinfo {author} {\bibfnamefont {Xingfei}\ \bibnamefont
  {Wei}}\ and\ \bibinfo {author} {\bibfnamefont {Tengfei}\ \bibnamefont
  {Luo}},\ }\bibfield  {title} {\enquote {\bibinfo {title} {Effects of
  electrostatic interaction and chirality on the friction coefficient of water
  flow inside single-walled carbon nanotubes and boron nitride nanotubes},}\
  }\href {\doibase 10.1021/acs.jpcc.7b11657} {\bibfield  {journal} {\bibinfo
  {journal} {J. Phys. Chem. C}\ }\textbf {\bibinfo {volume} {122}},\ \bibinfo
  {pages} {5131--5140} (\bibinfo {year} {2018})}\BibitemShut {NoStop}%
\bibitem [{\citenamefont {Sam}\ \emph {et~al.}(2019)\citenamefont {Sam},
  \citenamefont {{Vishnu Prasad}},\ and\ \citenamefont {Sathian}}]{Sam2019}%
  \BibitemOpen
  \bibfield  {author} {\bibinfo {author} {\bibfnamefont {Alan}\ \bibnamefont
  {Sam}}, \bibinfo {author} {\bibfnamefont {K.}~\bibnamefont {{Vishnu
  Prasad}}}, \ and\ \bibinfo {author} {\bibfnamefont {Sarith~P.}\ \bibnamefont
  {Sathian}},\ }\bibfield  {title} {\enquote {\bibinfo {title} {Water flow in
  carbon nanotubes: The role of tube chirality},}\ }\href {\doibase
  10.1039/c9cp00429g} {\bibfield  {journal} {\bibinfo  {journal} {Phys. Chem.
  Chem. Phys.}\ }\textbf {\bibinfo {volume} {21}},\ \bibinfo {pages}
  {6566--6573} (\bibinfo {year} {2019})}\BibitemShut {NoStop}%
\bibitem [{\citenamefont {Jorgensen}\ \emph {et~al.}(1983)\citenamefont
  {Jorgensen}, \citenamefont {Chandrasekhar}, \citenamefont {Madura},
  \citenamefont {Impey},\ and\ \citenamefont {Klein}}]{Jorgensen83}%
  \BibitemOpen
  \bibfield  {author} {\bibinfo {author} {\bibfnamefont {William~L.}\
  \bibnamefont {Jorgensen}}, \bibinfo {author} {\bibfnamefont {Jayaraman}\
  \bibnamefont {Chandrasekhar}}, \bibinfo {author} {\bibfnamefont {Jeffry~D.}\
  \bibnamefont {Madura}}, \bibinfo {author} {\bibfnamefont {Roger~W.}\
  \bibnamefont {Impey}}, \ and\ \bibinfo {author} {\bibfnamefont {Michael~L.}\
  \bibnamefont {Klein}},\ }\bibfield  {title} {\enquote {\bibinfo {title}
  {Comparison of simple potential functions for simulating liquid water},}\
  }\href {\doibase 10.1063/1.445869} {\bibfield  {journal} {\bibinfo  {journal}
  {J. Chem. Phys.}\ }\textbf {\bibinfo {volume} {79}},\ \bibinfo {pages}
  {926--935} (\bibinfo {year} {1983})}\BibitemShut {NoStop}%
\bibitem [{\citenamefont {Bondi}(1964)}]{Bondi1964}%
  \BibitemOpen
  \bibfield  {author} {\bibinfo {author} {\bibfnamefont {A.}~\bibnamefont
  {Bondi}},\ }\bibfield  {title} {\enquote {\bibinfo {title} {van der {Waals}
  volumes and radii},}\ }\href {\doibase 10.1021/j100785a001} {\bibfield
  {journal} {\bibinfo  {journal} {J. Phys. Chem.}\ }\textbf {\bibinfo {volume}
  {68}},\ \bibinfo {pages} {441--451} (\bibinfo {year} {1964})}\BibitemShut
  {NoStop}%
\bibitem [{\citenamefont {Mantina}\ \emph {et~al.}(2009)\citenamefont
  {Mantina}, \citenamefont {Chamberlin}, \citenamefont {Valero}, \citenamefont
  {Cramer},\ and\ \citenamefont {Truhlar}}]{Mantina2009}%
  \BibitemOpen
  \bibfield  {author} {\bibinfo {author} {\bibfnamefont {Manjeera}\
  \bibnamefont {Mantina}}, \bibinfo {author} {\bibfnamefont {Adam~C.}\
  \bibnamefont {Chamberlin}}, \bibinfo {author} {\bibfnamefont {Rosendo}\
  \bibnamefont {Valero}}, \bibinfo {author} {\bibfnamefont {Christopher~J.}\
  \bibnamefont {Cramer}}, \ and\ \bibinfo {author} {\bibfnamefont {Donald~G.}\
  \bibnamefont {Truhlar}},\ }\bibfield  {title} {\enquote {\bibinfo {title}
  {Consistent van der {W}aals radii for the whole main group},}\ }\href
  {\doibase 10.1021/jp8111556} {\bibfield  {journal} {\bibinfo  {journal} {J.
  Phys. Chem. A}\ }\textbf {\bibinfo {volume} {113}},\ \bibinfo {pages}
  {5806--5812} (\bibinfo {year} {2009})}\BibitemShut {NoStop}%
\bibitem [{\citenamefont {CY}\ and\ \citenamefont {NR}()}]{Aluru2007}%
  \BibitemOpen
  \bibfield  {author} {\bibinfo {author} {\bibfnamefont {Won}\ \bibnamefont
  {CY}}\ and\ \bibinfo {author} {\bibfnamefont {Aluru}\ \bibnamefont {NR}},\
  }\bibfield  {title} {\enquote {\bibinfo {title} {Water permeation through a
  subnanometer boron nitride nanotube},}\ }\href@noop {} {\bibfield  {journal}
  {\bibinfo  {journal} {J. Am. Chem. Soc.}\ }\textbf {\bibinfo {volume}
  {129}}}\BibitemShut {NoStop}%
\bibitem [{\citenamefont {Suk}\ \emph {et~al.}(2008)\citenamefont {Suk},
  \citenamefont {Raghunathan},\ and\ \citenamefont {Aluru}}]{Suk2008}%
  \BibitemOpen
  \bibfield  {author} {\bibinfo {author} {\bibfnamefont {M.~E.}\ \bibnamefont
  {Suk}}, \bibinfo {author} {\bibfnamefont {A.~V.}\ \bibnamefont
  {Raghunathan}}, \ and\ \bibinfo {author} {\bibfnamefont {N.~R.}\ \bibnamefont
  {Aluru}},\ }\bibfield  {title} {\enquote {\bibinfo {title} {Fast reverse
  osmosis using boron nitride and carbon nanotubes},}\ }\href {\doibase
  10.1063/1.2907333} {\bibfield  {journal} {\bibinfo  {journal} {Appl. Phys.
  Lett.}\ }\textbf {\bibinfo {volume} {92}},\ \bibinfo {pages} {133120}
  (\bibinfo {year} {2008})}\BibitemShut {NoStop}%
\bibitem [{\citenamefont {Ritos}\ \emph {et~al.}(2014)\citenamefont {Ritos},
  \citenamefont {Mattia}, \citenamefont {Calabr{\`{o}}},\ and\ \citenamefont
  {Reese}}]{Ritos2014}%
  \BibitemOpen
  \bibfield  {author} {\bibinfo {author} {\bibfnamefont {Konstantinos}\
  \bibnamefont {Ritos}}, \bibinfo {author} {\bibfnamefont {Davide}\
  \bibnamefont {Mattia}}, \bibinfo {author} {\bibfnamefont {Francesco}\
  \bibnamefont {Calabr{\`{o}}}}, \ and\ \bibinfo {author} {\bibfnamefont
  {Jason~M.}\ \bibnamefont {Reese}},\ }\bibfield  {title} {\enquote {\bibinfo
  {title} {Flow enhancement in nanotubes of different materials and lengths},}\
  }\href {\doibase 10.1063/1.4846300} {\bibfield  {journal} {\bibinfo
  {journal} {J. Chem. Phys.}\ }\textbf {\bibinfo {volume} {140}},\ \bibinfo
  {pages} {014702} (\bibinfo {year} {2014})}\BibitemShut {NoStop}%
\bibitem [{\citenamefont {Secchi}\ \emph {et~al.}(2016)\citenamefont {Secchi},
  \citenamefont {Marbach}, \citenamefont {Nigu\`{e}s}, \citenamefont {Stein},
  \citenamefont {Siria},\ and\ \citenamefont {Bocquet}}]{Secchi2016}%
  \BibitemOpen
  \bibfield  {author} {\bibinfo {author} {\bibfnamefont {Eleonora}\
  \bibnamefont {Secchi}}, \bibinfo {author} {\bibfnamefont {Sophie}\
  \bibnamefont {Marbach}}, \bibinfo {author} {\bibfnamefont {Antoine}\
  \bibnamefont {Nigu\`{e}s}}, \bibinfo {author} {\bibfnamefont {Derek}\
  \bibnamefont {Stein}}, \bibinfo {author} {\bibfnamefont {Alessandro}\
  \bibnamefont {Siria}}, \ and\ \bibinfo {author} {\bibfnamefont {Lyd\'{e}ric}\
  \bibnamefont {Bocquet}},\ }\bibfield  {title} {\enquote {\bibinfo {title}
  {Massive radius-dependent flow slippage in carbon nanotubes},}\ }\href
  {\doibase 10.1038/nature19315} {\bibfield  {journal} {\bibinfo  {journal}
  {Nature}\ }\textbf {\bibinfo {volume} {537}},\ \bibinfo {pages} {210--213}
  (\bibinfo {year} {2016})}\BibitemShut {NoStop}%
\bibitem [{\citenamefont {Thomas}\ \emph {et~al.}(2014)\citenamefont {Thomas},
  \citenamefont {Corry},\ and\ \citenamefont {Hilder}}]{Thomas2014}%
  \BibitemOpen
  \bibfield  {author} {\bibinfo {author} {\bibfnamefont {Michael}\ \bibnamefont
  {Thomas}}, \bibinfo {author} {\bibfnamefont {Ben}\ \bibnamefont {Corry}}, \
  and\ \bibinfo {author} {\bibfnamefont {Tamsyn~A.}\ \bibnamefont {Hilder}},\
  }\bibfield  {title} {\enquote {\bibinfo {title} {What have we learnt about
  the mechanisms of rapid water transport, ion rejection and selectivity in
  nanopores from molecular simulation?}}\ }\href {\doibase
  10.1002/smll.201302968} {\bibfield  {journal} {\bibinfo  {journal} {Small}\
  }\textbf {\bibinfo {volume} {10}},\ \bibinfo {pages} {1453--1465} (\bibinfo
  {year} {2014})}\BibitemShut {NoStop}%
\bibitem [{\citenamefont {Hilder}\ \emph {et~al.}(2010)\citenamefont {Hilder},
  \citenamefont {Yang}, \citenamefont {Ganesh}, \citenamefont {Gordon},
  \citenamefont {Bliznyuk}, \citenamefont {Rendell},\ and\ \citenamefont
  {Chung}}]{Hilder2010}%
  \BibitemOpen
  \bibfield  {author} {\bibinfo {author} {\bibfnamefont {T.A.}\ \bibnamefont
  {Hilder}}, \bibinfo {author} {\bibfnamefont {R.}~\bibnamefont {Yang}},
  \bibinfo {author} {\bibfnamefont {V.}~\bibnamefont {Ganesh}}, \bibinfo
  {author} {\bibfnamefont {D.}~\bibnamefont {Gordon}}, \bibinfo {author}
  {\bibfnamefont {A.}~\bibnamefont {Bliznyuk}}, \bibinfo {author}
  {\bibfnamefont {A.P.}\ \bibnamefont {Rendell}}, \ and\ \bibinfo {author}
  {\bibfnamefont {S.-H.}\ \bibnamefont {Chung}},\ }\bibfield  {title} {\enquote
  {\bibinfo {title} {Validity of current force fields for simulations on boron
  nitride nanotubes},}\ }\href {\doibase 10.1049/mnl.2009.0112} {\bibfield
  {journal} {\bibinfo  {journal} {Micro {\&} Nano Lett.}\ }\textbf {\bibinfo
  {volume} {5}},\ \bibinfo {pages} {150} (\bibinfo {year} {2010})}\BibitemShut
  {NoStop}%
\bibitem [{\citenamefont {Melillo}\ \emph {et~al.}(2011)\citenamefont
  {Melillo}, \citenamefont {Zhu}, \citenamefont {Snyder},\ and\ \citenamefont
  {Mittal}}]{Melillo2011}%
  \BibitemOpen
  \bibfield  {author} {\bibinfo {author} {\bibfnamefont {Matthew}\ \bibnamefont
  {Melillo}}, \bibinfo {author} {\bibfnamefont {Fangqiang}\ \bibnamefont
  {Zhu}}, \bibinfo {author} {\bibfnamefont {Mark~A.}\ \bibnamefont {Snyder}}, \
  and\ \bibinfo {author} {\bibfnamefont {Jeetain}\ \bibnamefont {Mittal}},\
  }\bibfield  {title} {\enquote {\bibinfo {title} {Water transport through
  nanotubes with varying interaction strength between tube wall and water},}\
  }\href {\doibase 10.1021/jz2012319} {\bibfield  {journal} {\bibinfo
  {journal} {J. Phys. Chem. Lett.}\ }\textbf {\bibinfo {volume} {2}},\ \bibinfo
  {pages} {2978--2983} (\bibinfo {year} {2011})}\BibitemShut {NoStop}%
\bibitem [{\citenamefont {Tocci}\ \emph {et~al.}(2014)\citenamefont {Tocci},
  \citenamefont {Joly},\ and\ \citenamefont {Michaelides}}]{Tocci2014}%
  \BibitemOpen
  \bibfield  {author} {\bibinfo {author} {\bibfnamefont {Gabriele}\
  \bibnamefont {Tocci}}, \bibinfo {author} {\bibfnamefont {Laurent}\
  \bibnamefont {Joly}}, \ and\ \bibinfo {author} {\bibfnamefont {Angelos}\
  \bibnamefont {Michaelides}},\ }\bibfield  {title} {\enquote {\bibinfo {title}
  {Friction of water on graphene and hexagonal boron nitride from ab initio
  methods: Very different slippage despite very similar interface
  structures},}\ }\href {\doibase 10.1021/nl502837d} {\bibfield  {journal}
  {\bibinfo  {journal} {Nano Lett.}\ }\textbf {\bibinfo {volume} {14}},\
  \bibinfo {pages} {6872--6877} (\bibinfo {year} {2014})}\BibitemShut {NoStop}%
\bibitem [{\citenamefont {Gelb}\ \emph {et~al.}(1999)\citenamefont {Gelb},
  \citenamefont {Gubbins}, \citenamefont {Radhakrishnan},\ and\ \citenamefont
  {Sliwinska-Bartkowiak}}]{Gelb1999}%
  \BibitemOpen
  \bibfield  {author} {\bibinfo {author} {\bibfnamefont {Lev~D.}\ \bibnamefont
  {Gelb}}, \bibinfo {author} {\bibfnamefont {K.~E.}\ \bibnamefont {Gubbins}},
  \bibinfo {author} {\bibfnamefont {R.}~\bibnamefont {Radhakrishnan}}, \ and\
  \bibinfo {author} {\bibfnamefont {M.}~\bibnamefont {Sliwinska-Bartkowiak}},\
  }\bibfield  {title} {\enquote {\bibinfo {title} {Phase separation in confined
  systems},}\ }\href {\doibase 10.1088/0034-4885/62/12/201} {\bibfield
  {journal} {\bibinfo  {journal} {Rep. Prog. Phys.}\ }\textbf {\bibinfo
  {volume} {62}},\ \bibinfo {pages} {1573--1659} (\bibinfo {year}
  {1999})}\BibitemShut {NoStop}%
\bibitem [{\citenamefont {Czwartos}\ \emph {et~al.}(2005)\citenamefont
  {Czwartos}, \citenamefont {Coasne}, \citenamefont {{Gubbins *}},
  \citenamefont {Hung},\ and\ \citenamefont
  {Sliwinska-Bartkowiak}}]{Czwartos2005}%
  \BibitemOpen
  \bibfield  {author} {\bibinfo {author} {\bibfnamefont {J.}~\bibnamefont
  {Czwartos}}, \bibinfo {author} {\bibfnamefont {B.}~\bibnamefont {Coasne}},
  \bibinfo {author} {\bibfnamefont {K.~E.}\ \bibnamefont {{Gubbins *}}},
  \bibinfo {author} {\bibfnamefont {F.~R.}\ \bibnamefont {Hung}}, \ and\
  \bibinfo {author} {\bibfnamefont {M.}~\bibnamefont {Sliwinska-Bartkowiak}},\
  }\bibfield  {title} {\enquote {\bibinfo {title} {Freezing and melting of
  azeotropic mixtures confined in nanopores: experiment and molecular
  simulation},}\ }\href {\doibase 10.1080/00268970500200101} {\bibfield
  {journal} {\bibinfo  {journal} {Mol. Phys.}\ }\textbf {\bibinfo {volume}
  {103}},\ \bibinfo {pages} {3103--3113} (\bibinfo {year} {2005})}\BibitemShut
  {NoStop}%
\bibitem [{\citenamefont {Srivastava}\ \emph {et~al.}(2004)\citenamefont
  {Srivastava}, \citenamefont {Srivastava}, \citenamefont {Talapatra},
  \citenamefont {Vajtai},\ and\ \citenamefont {Ajayan}}]{Srivastava2004}%
  \BibitemOpen
  \bibfield  {author} {\bibinfo {author} {\bibfnamefont {A.}~\bibnamefont
  {Srivastava}}, \bibinfo {author} {\bibfnamefont {O.~N.}\ \bibnamefont
  {Srivastava}}, \bibinfo {author} {\bibfnamefont {S.}~\bibnamefont
  {Talapatra}}, \bibinfo {author} {\bibfnamefont {R.}~\bibnamefont {Vajtai}}, \
  and\ \bibinfo {author} {\bibfnamefont {P.~M.}\ \bibnamefont {Ajayan}},\
  }\bibfield  {title} {\enquote {\bibinfo {title} {Carbon nanotube filters},}\
  }\href {\doibase 10.1038/nmat1192} {\bibfield  {journal} {\bibinfo  {journal}
  {Nat. Mater.}\ }\textbf {\bibinfo {volume} {3}},\ \bibinfo {pages} {610--614}
  (\bibinfo {year} {2004})}\BibitemShut {NoStop}%
\bibitem [{H2O(1986)}]{H2O-CRC86}%
  \BibitemOpen
  \href@noop {} {\emph {\bibinfo {title} {{CRC} {Handbook} of {Chemistry} and
  {Physics}}}},\ \bibinfo {edition} {67th}\ ed.\ (\bibinfo  {publisher} {CRC
  Press},\ \bibinfo {address} {Boca Raton, FL, USA},\ \bibinfo {year}
  {1986})\BibitemShut {NoStop}%
\bibitem [{\citenamefont {Joseph}\ and\ \citenamefont
  {Aluru}(2008)}]{Joseph2008}%
  \BibitemOpen
  \bibfield  {author} {\bibinfo {author} {\bibfnamefont {Sony}\ \bibnamefont
  {Joseph}}\ and\ \bibinfo {author} {\bibfnamefont {N.~R.}\ \bibnamefont
  {Aluru}},\ }\bibfield  {title} {\enquote {\bibinfo {title} {Why are carbon
  nanotubes fast transporters of water?}}\ }\href {\doibase 10.1021/nl072385q}
  {\bibfield  {journal} {\bibinfo  {journal} {Nano Lett.}\ }\textbf {\bibinfo
  {volume} {8}},\ \bibinfo {pages} {452--458} (\bibinfo {year}
  {2008})}\BibitemShut {NoStop}%
\bibitem [{\citenamefont {Hanasaki}\ and\ \citenamefont
  {Nakatani}(2006)}]{Hanasaki2006}%
  \BibitemOpen
  \bibfield  {author} {\bibinfo {author} {\bibfnamefont {Itsuo}\ \bibnamefont
  {Hanasaki}}\ and\ \bibinfo {author} {\bibfnamefont {Akihiro}\ \bibnamefont
  {Nakatani}},\ }\bibfield  {title} {\enquote {\bibinfo {title} {Flow structure
  of water in carbon nanotubes: Poiseuille type or plug-like?}}\ }\href
  {\doibase 10.1063/1.2187971} {\bibfield  {journal} {\bibinfo  {journal} {J.
  Chem. Phys.}\ }\textbf {\bibinfo {volume} {124}},\ \bibinfo {pages} {144708}
  (\bibinfo {year} {2006})}\BibitemShut {NoStop}%
\bibitem [{\citenamefont {Joly}(2011)}]{Joly2011}%
  \BibitemOpen
  \bibfield  {author} {\bibinfo {author} {\bibfnamefont {Laurent}\ \bibnamefont
  {Joly}},\ }\bibfield  {title} {\enquote {\bibinfo {title} {Capillary filling
  with giant liquid/solid slip: Dynamics of water uptake by carbon
  nanotubes},}\ }\href {\doibase 10.1063/1.3664622} {\bibfield  {journal}
  {\bibinfo  {journal} {J. Chem. Phys.}\ }\textbf {\bibinfo {volume} {135}},\
  \bibinfo {pages} {214705} (\bibinfo {year} {2011})}\BibitemShut {NoStop}%
\bibitem [{\citenamefont {Sam}\ \emph {et~al.}(2017)\citenamefont {Sam},
  \citenamefont {Kannam}, \citenamefont {Hartkamp},\ and\ \citenamefont
  {Sathian}}]{Sam2017}%
  \BibitemOpen
  \bibfield  {author} {\bibinfo {author} {\bibfnamefont {Alan}\ \bibnamefont
  {Sam}}, \bibinfo {author} {\bibfnamefont {Sridhar~Kumar}\ \bibnamefont
  {Kannam}}, \bibinfo {author} {\bibfnamefont {Remco}\ \bibnamefont
  {Hartkamp}}, \ and\ \bibinfo {author} {\bibfnamefont {Sarith~P.}\
  \bibnamefont {Sathian}},\ }\bibfield  {title} {\enquote {\bibinfo {title}
  {Water flow in carbon nanotubes: The effect of tube flexibility and
  thermostat},}\ }\href {\doibase 10.1063/1.4985252} {\bibfield  {journal}
  {\bibinfo  {journal} {J. Chem. Phys.}\ }\textbf {\bibinfo {volume} {146}}
  (\bibinfo {year} {2017}),\ 10.1063/1.4985252}\BibitemShut {NoStop}%
\bibitem [{\citenamefont {Tao}\ \emph {et~al.}(2018)\citenamefont {Tao},
  \citenamefont {Song}, \citenamefont {Zhao}, \citenamefont {Zhao},\ and\
  \citenamefont {Liu}}]{Tao2018}%
  \BibitemOpen
  \bibfield  {author} {\bibinfo {author} {\bibfnamefont {Jiabo}\ \bibnamefont
  {Tao}}, \bibinfo {author} {\bibfnamefont {Xianyu}\ \bibnamefont {Song}},
  \bibinfo {author} {\bibfnamefont {Teng}\ \bibnamefont {Zhao}}, \bibinfo
  {author} {\bibfnamefont {Shuangliang}\ \bibnamefont {Zhao}}, \ and\ \bibinfo
  {author} {\bibfnamefont {Honglai}\ \bibnamefont {Liu}},\ }\bibfield  {title}
  {\enquote {\bibinfo {title} {Confinement effect on water transport in cnt
  membranes},}\ }\href {\doibase 10.1016/j.ces.2018.05.018} {\bibfield
  {journal} {\bibinfo  {journal} {Chem. Eng. Sci.}\ }\textbf {\bibinfo {volume}
  {192}},\ \bibinfo {pages} {1252--1259} (\bibinfo {year} {2018})}\BibitemShut
  {NoStop}%
\bibitem [{\citenamefont {Jassby}\ \emph {et~al.}(2018)\citenamefont {Jassby},
  \citenamefont {Cath},\ and\ \citenamefont {Buisson}}]{Jassby2018}%
  \BibitemOpen
  \bibfield  {author} {\bibinfo {author} {\bibfnamefont {David}\ \bibnamefont
  {Jassby}}, \bibinfo {author} {\bibfnamefont {Tzahi~Y.}\ \bibnamefont {Cath}},
  \ and\ \bibinfo {author} {\bibfnamefont {Herve}\ \bibnamefont {Buisson}},\
  }\bibfield  {title} {\enquote {\bibinfo {title} {The role of nanotechnology
  in industrial water treatment},}\ }\href {\doibase 10.1038/s41565-018-0234-8}
  {\bibfield  {journal} {\bibinfo  {journal} {Nat. Nanotechnol.}\ }\textbf
  {\bibinfo {volume} {13}},\ \bibinfo {pages} {670--672} (\bibinfo {year}
  {2018})}\BibitemShut {NoStop}%
\bibitem [{\citenamefont {Albergamo}\ \emph {et~al.}(2018)\citenamefont
  {Albergamo}, \citenamefont {Blankert}, \citenamefont {Cornelissen},
  \citenamefont {Hofs}, \citenamefont {Knibbe}, \citenamefont {van~der Meer},\
  and\ \citenamefont {de~Voogt}}]{Albergamo2018}%
  \BibitemOpen
  \bibfield  {author} {\bibinfo {author} {\bibfnamefont {Vittorio}\
  \bibnamefont {Albergamo}}, \bibinfo {author} {\bibfnamefont {Bastiaan}\
  \bibnamefont {Blankert}}, \bibinfo {author} {\bibfnamefont {Emile}\
  \bibnamefont {Cornelissen}}, \bibinfo {author} {\bibfnamefont {Bas}\
  \bibnamefont {Hofs}}, \bibinfo {author} {\bibfnamefont {Willem-Jan}\
  \bibnamefont {Knibbe}}, \bibinfo {author} {\bibfnamefont {Walter}\
  \bibnamefont {van~der Meer}}, \ and\ \bibinfo {author} {\bibfnamefont {Pim}\
  \bibnamefont {de~Voogt}},\ }\bibfield  {title} {\enquote {\bibinfo {title}
  {Removal of polar organic micropollutants by pilot-scale reverse osmosis
  drinking water treatment},}\ }\href {\doibase 10.1016/j.watres.2018.09.029}
  {\bibfield  {journal} {\bibinfo  {journal} {Wat. Res.}\ }\textbf {\bibinfo
  {volume} {148}},\ \bibinfo {pages} {535--545} (\bibinfo {year}
  {2018})}\BibitemShut {NoStop}%
\bibitem [{\citenamefont {Werber}\ \emph
  {et~al.}(2016{\natexlab{a}})\citenamefont {Werber}, \citenamefont {Osuji},\
  and\ \citenamefont {Elimelech}}]{Werber2016}%
  \BibitemOpen
  \bibfield  {author} {\bibinfo {author} {\bibfnamefont {Jay~R.}\ \bibnamefont
  {Werber}}, \bibinfo {author} {\bibfnamefont {Chinedum~O.}\ \bibnamefont
  {Osuji}}, \ and\ \bibinfo {author} {\bibfnamefont {Menachem}\ \bibnamefont
  {Elimelech}},\ }\bibfield  {title} {\enquote {\bibinfo {title} {Materials for
  next-generation desalination and water purification membranes},}\ }\href
  {\doibase 10.1038/natrevmats.2016.18} {\bibfield  {journal} {\bibinfo
  {journal} {Nat. Rev. Mater.}\ }\textbf {\bibinfo {volume} {1}},\ \bibinfo
  {pages} {16018} (\bibinfo {year} {2016}{\natexlab{a}})}\BibitemShut {NoStop}%
\bibitem [{\citenamefont {Werber}\ \emph
  {et~al.}(2016{\natexlab{b}})\citenamefont {Werber}, \citenamefont
  {Deshmukh},\ and\ \citenamefont {Elimelech}}]{Werber2016a}%
  \BibitemOpen
  \bibfield  {author} {\bibinfo {author} {\bibfnamefont {Jay~R.}\ \bibnamefont
  {Werber}}, \bibinfo {author} {\bibfnamefont {Akshay}\ \bibnamefont
  {Deshmukh}}, \ and\ \bibinfo {author} {\bibfnamefont {Menachem}\ \bibnamefont
  {Elimelech}},\ }\bibfield  {title} {\enquote {\bibinfo {title} {The critical
  need for increased selectivity, not increased water permeability, for
  desalination membranes},}\ }\href {\doibase 10.1021/acs.estlett.6b00050}
  {\bibfield  {journal} {\bibinfo  {journal} {Environ. Sci. Tech. Let.}\
  }\textbf {\bibinfo {volume} {3}},\ \bibinfo {pages} {112--120} (\bibinfo
  {year} {2016}{\natexlab{b}})}\BibitemShut {NoStop}%
\bibitem [{\citenamefont {Elimelech}\ and\ \citenamefont
  {Phillip}(2011)}]{Elimelech2011}%
  \BibitemOpen
  \bibfield  {author} {\bibinfo {author} {\bibfnamefont {Menachem}\
  \bibnamefont {Elimelech}}\ and\ \bibinfo {author} {\bibfnamefont {William~A}\
  \bibnamefont {Phillip}},\ }\bibfield  {title} {\enquote {\bibinfo {title}
  {The future of seawater desalination: Energy, technology, and the
  environment},}\ }\href {\doibase 10.1126/science.1200488} {\bibfield
  {journal} {\bibinfo  {journal} {Science}\ }\textbf {\bibinfo {volume}
  {333}},\ \bibinfo {pages} {712--717} (\bibinfo {year} {2011})}\BibitemShut
  {NoStop}%
\end{thebibliography}

%

\end{document}